\documentclass{article}
\usepackage{amsfonts,amssymb,amsmath,graphicx}
\usepackage{theorem}
\usepackage{caption}
\usepackage{cite}
\usepackage{url}
\usepackage{enumerate}
\usepackage{tikz}
\usepackage[margin=1.2in]{geometry}

\usetikzlibrary{arrows}
\tikzstyle{block}=[draw opacity=0.7,line width=1.4cm]
\usetikzlibrary{positioning,arrows,chains,matrix,scopes,fit,decorations.markings,decorations,decorations.pathreplacing}
\tikzset{
  on each segment/.style={
    decorate,
    decoration={
      show path construction,
      moveto code={},
      lineto code={
        \path [#1]
        (\tikzinputsegmentfirst) -- (\tikzinputsegmentlast);
      },
      curveto code={
        \path [#1] (\tikzinputsegmentfirst)
        .. controls
        (\tikzinputsegmentsupporta) and (\tikzinputsegmentsupportb)
        ..
        (\tikzinputsegmentlast);
      },
      closepath code={
        \path [#1]
        (\tikzinputsegmentfirst) -- (\tikzinputsegmentlast);
      },
    },
  },
  mid arrow/.style={postaction={decorate,decoration={
        markings,
        mark=at position .4 with {\arrow[#1]{stealth}}
      }}},
}

\newtheorem{definition}{Definition}[section]

\newtheorem{theorem}[definition]{Theorem}
\newtheorem{corollary}[definition]{Corollary}
\newtheorem{lemma}[definition]{Lemma}
\newtheorem{rmk}[definition]{Remark}
\numberwithin{equation}{section}

\DeclareMathOperator{\tr}{tr}

\newcommand{\beq}{\begin{equation}}
\newcommand{\eeq}{\end{equation}}
\newcommand{\bea}{\begin{eqnarray}}
\newcommand{\eea}{\end{eqnarray}}
\newcommand{\beano}{\begin{eqnarray*}}
\newcommand{\eeano}{\end{eqnarray*}}
\newcommand{\bma}{\begin{pmatrix}}
\newcommand{\ema}{\end{pmatrix}}

\def\cA{{\cal A}}      \def\cB{{\cal B}}      \def\cC{{\cal C}}
\def\cD{{\cal D}}            
      \def\cH{{\cal H}}      
      \def\cK{{\cal K}}      \def\cL{{\cal L}}
\def\cM{{\cal M}}      \def\cN{{\cal N}}      
\def\cP{{\cal P}}      \def\cQ{{\cal Q}}      \def\cR{{\cal R}}
\def\cS{{\cal S}}      \def\cT{{\cal T}}      \def\cU{{\cal U}}
\def\cV{{\cal V}}

\def\fD{{\mathfrak D}}

\def\fN{{\mathfrak N}}

\newcommand{\CC}{{\mathbb C}}

\newcommand{\RR}{\mbox{${\mathbb R}$}}

\newcommand{\ZZ}{{\mathbb Z}}

\newcommand{\prf}{\underline{Proof:}\ }
\newcommand{\finprf}{\null \hfill {\rule{5pt}{5pt}}\indent}
\newcommand{\ie}{{\it ie}.\ }

\title{On an inverse scattering transform  \\ for the nonlinear Sch\"odinger equation on the half-line\footnote{This project is support by NSFC No.~11601312, 11875040.  }}
\date{\empty}
\author{ Cheng Zhang\footnote{Email address: ch.zhang.maths@gmail.com} 
  \\ \\
  \sc \small Department of Mathematics \\
  \sc  \small Shanghai University \\
  \sc  \small Shanghai, 200444, China \\ \\
}

\begin{document}
\maketitle

\begin{abstract}
In this paper, we develop an inverse scattering transform for the integrable focusing nonlinear Schr\"odinger (NLS) equation on the half-line subject to  a class of boundary conditions. The method is based on the notions of integrable boundary conditions and  double-row monodromy matrix, developed by Sklyanin, which characterize initial-boundary value problems for  NLS on an interval. It follows from Sklyanin's approach that a hierarchy of integrable boundary conditions can be derived based on a semi-discrete type Lax pair involving the time-part Lax matrix of NLS and a reflection matrix. The inverse scattering transform relies on the formulation of a scattering system for the double-row monodromy matrix characterizing an interval problem. The scattering system for the half-line problem for NLS is obtained by extending one endpoint of the interval to infinity. Then, we derive spectral and analytic properties of the scattering systems, and set up the inverse part using a Riemann-Hilbert formulation. We also show that our approach is equivalent to a nonlinear method of reflection by extending the initial-boundary value problem on the half-line to an initial value on the whole axis. Explicit examples of soliton solutions on the half-line are provided. Although we only consider the NLS model as a particular example, the method we present in this paper can  be readily applied to a wide range of integrable PDEs on the half-line.  
\vspace{.2cm}


\vspace{.2cm}

\end{abstract}

\section{Introduction}

The inverse scattering transform is  a well-established analytic approach to dealing with initial value problems for a wide class of two-dimensional integrable nonlinear partial differential equations (PDEs) \cite{kdv, ZS2,AKNS}. The crucial aspect of successively applying this method  depends  on the asymptotic behaviors and/or boundary conditions of the model under consideration. Take the focusing nonlinear Schr\"odinger (NLS) equation as our primary example. To integrate the NLS equation on the whole axis using the inverse scattering transform, the vanishing boundary conditions (the NLS fields vanish rapidly as the space variable tends to infinity) should be imposed \cite{faddeev, ZS2}. In
the Hamiltonian formulation of integrable field theories, the vanishing boundary conditions are also needed to ensure the existence of infinitely many conserved quantities in involution \cite{G1, faddeev,f1}.
Indeed, one could argue that an integrable PDE is said to be integrable  only if it is subject to a special class of  boundary behaviors. For the NLS equation, the typical choices are either the vanishing boundary conditions for the space variable defined on the whole axis,  or the periodic boundary conditions for the space variable defined on a circle. The former case could be understood as the latter case with the length of the periodicity tending  to infinity \cite{faddeev}. This aspect is well summarized  by Krichever and Novikov in their review paper \cite{KN1}:
\begin{quotation}
\it  ``Let us point out that the KdV system, as well as other nontrivial completely integrable by IST (inverse scattering transform) partial differential equation
(PDE) systems, are indeed completely integrable in any reasonable sense for rapidly decreasing
or periodic (quasi-periodic) boundary conditions only.''
\end{quotation}

In contrast to periodic problems (and  problems on the whole axis as limiting cases),  {\em open boundary problems}  for integrable PDEs deal with initial-boundary value problems where the space variable is defined either on a closed interval or on the half-line. One of the systematic approaches to selecting admissible class of boundary conditions, known as {\em integrable boundary conditions} that preserve the integrability of the model in the presence of boundaries or a boundary,  was due to Sklyanin  in the framework of integrable Hamiltonian field theories  \cite{SKBC} (see also  \cite{ACC} for recent developments) and their quantum analogues \cite{Cherednik1, sklyanin1988boundary}. In Sklyanin's formalism, the central object is the so-called {\em double-row monodromy matrix}, which is constructed with the help of {\em reflection matrices} as solutions of the {\em classical reflection equations}.
For an integrable PDE defined on an interval, the boundary conditions at the two endpoints are encoded into certain refection matrices. The integrability is ensured by the existence of infinitely many conserved quantities in involution. Since Sklyanin's original work \cite{SKBC}, it has been known that a wide class of integrable PDEs, including the NLS, Sine-Gordon and Landau-Lifshitz equations as typical examples, is integrable on an interval equipped with integrable boundary conditions. 
In spite of a good understanding of the integrable aspects of these models, little work in line with Sklyanin's formalism has been known to address the analytic approaches to solving the associated initial-boundary value problems.

The Fokas' unified transform has been known to be a generic analytic approach to dealing with    initial-boundary value problems for a wide class of integrable PDEs \cite{fokas1997unified, fokas2002integrable, fokas2008}. It can be considered as a generalization of the inverse scattering transform. The key idea of the unified transform method is to treat simultaneously the initial and boundary data in the direct scattering process. Hence, both the initial and boundary data are treated  at the same footing. The associated scattering functions can be put into certain functional constraints  formulated as Riemann-Hilbert problems in the inverse space. Although Fokas' unified transform method has been successfully applied to a wide range of integrable PDEs with a boundary or boundaries, it is,  in general,  a difficult task to explicitly solve the Riemann-Hilbert problems. Moreover, in contrast to Sklyanin's approach, there is no clear definition of integrable boundary conditions in the unified transform method. Note that a special class of boundary conditions, called {\em linearizable} boundary conditions, do exist in Fokas' approach \cite{fokas2002integrable, fokas2008}. The linearizable boundary conditions reflect certain symmetry of the scattering functions, and can be used to reduce the Riemann-Hilbert problems. For certain models, they coincide with the integrable boundary conditions, {\it cf}.~\cite{SKBC, fokas2002integrable}. However, the linearizable boundary conditions are, a priori, not equivalent to integrable boundary conditions.  As an important application of the unified transform method, asymptotic solutions of integrable PDEs at large time, mostly accompanied with the linearizable boundary conditions, can be derived \cite{fokas2002integrable, fokas2008} using the nonlinear steepest descent method \cite{DP1}.  

The aim of the paper is to  develop an analytic method for solving a class of integrable  PDEs on the half-line equipped with integrable boundary conditions. We concentrate on the focusing NLS equation as our primary example. The NLS equation reads
\begin{equation}
  \label{eq:fnls}
  i q_t + q_{xx}+2  |q|^2 q= 0\,,\quad q:=q(x,t)\,.
\end{equation}
The  NLS field $q$  is a complex-valued function, and $x$ and $t$ are respectively the space and time variables. We use $q^*$ to denote the complex conjugate of $q$.  
The NLS equation is an important model in mathematical physics (see \cite{ZS2, faddeev, APT} for its integrable aspects), and possesses multi-soliton solutions as particular solutions. Initial value problems for NLS on the whole axis under the vanishing boundary conditions can be solved using the inverse scattering transform \cite{ZS2}.  

The NLS model on an interval or on the half-line has also been extensively studied in the literature. 
Sklyanin was the first to derive integrable boundary conditions, that are Robin boundary conditions, {\it cf}.~\eqref{eq:Rb1}, of the model defined on an interval. His work was  followed by other important contributions characterizing analytic and integrable aspects of the half-line NLS model  \cite{fokas1989initial, BT1, Tarasov, HH1}. In \cite{ZAMBON}, some new dynamical boundary conditions, {\it cf}.~\eqref{eq:iqt2}, were derived based on  Sklyanin's formalism and shown to be integrable.
The unified transform method was applied to the half-line NLS model, and  asymptotic soliton solutions at large time were derived using the nonlinear steepest descent method \cite{fokas2002integrable, fokas2005nonlinear}.  The qualitative behaviors of initial-boundary value problems for the NLS equation on the half-line under the homogeneous and non-homogeneous Robin boundary conditions can also be  found, for instance, in \cite{QBu, DP1
  , BSZ}. 

Soliton solutions of the NLS equation on the half-line subject to Robin boundary conditions were obtained rather recently in \cite{biondini2009solitons}  where a nonlinear method of reflection was implemented (see also \cite{fokas1989initial} for early developments). The method consists in extending the initial-boundary value problem on the half-line to an initial value problem on the whole axis. The extension reflects the space inversion symmetry of NLS \cite{Tarasov, HH1} and allows us to solve the model using the usual inverse scattering transform by uniquely looking at the positive semi axis. Generalization of this method can be found in \cite{CZ, bion2, CCAL}. 
Another direct approach, known as {\em dressing the boundary}, was introduced by the author to derive soliton solutions of NLS on the half-line \cite{ZC1}.  
This method is based on simultaneous dressing transformations for both the   {\em bulk} equation and the  boundary conditions, and  can be considered as a half-line version of the Darboux-dressing transformations. Applications of this method to the sine-Gordon and vector NLS models can be found in  \cite{ZCZ, ZCZ2}. In particular, soliton solutions of NLS on the half-line subject to the new dynamical boundary conditions obtained in \cite{ZAMBON}, {\it cf}.~\eqref{eq:iqt2},  were recently obtained in \cite{Gruner2, Xia} using this method.

Existence of soliton solutions for NLS (and other models) on the half-line subject to various type of integrable boundary conditions suggests that an analytic method (somehow ``missing'' in the literature) for solving integrable PDEs on the half-line with integrable boundary conditions in line with the tranditional inverse scattering transform for full line problems should exist. 
It also provide a strong evidence that  the solution structures of the half-line problems are essentially the same as those of the full line problems.

The main content of this paper is to develop the inverse scattering transform for integrable PDEs  on the half-line. 
This is based on the notions of integrable boundary conditions and double-row monodromy matrix introduced by Sklyanin \cite{SKBC, sklyanin1988boundary}. First, we will show that  Sklyanin’s formalism yields  a class of boundary conditions which belongs a hierarchy of integrable boundary conditions. This hierarchy is encoded into certain semi-discrete type Lax pair involving both the reflection matrix and the time-part Lax matrix of NLS. This provides new examples of integrable boundary conditions for NLS. Some of them are highly nonlinear, and can only be expressed in implicit forms, {\it cf}.~Case $5, 6$ in Section~\ref{sec.32}.  
The analytic method is based on a formulation of the scattering system in the space of the spectral parameter for the double-row monodromy
matrix, which characterizes an initial-boundary value problem for NLS on an interval. The scattering system for a half-line problem  is obtained by sending one endpoint of the interval to infinity. This allows us to identify half-line versions of Jost solutions. By deriving the associated spectral and analytic properties of the scattering systems, one can set up the inverse scattering transform  by virtue of a Riemann-Hilbert formulation. Using a systematic extension scheme, we also show that our approach is equivalent to the nonlinear method of reflection \cite{fokas1989initial, biondini2009solitons}. As particular applications, explicit examples of NLS solitons on the half-line subject to various boundary conditions belonging to the hierarchy  are provided. Although we only consider the NLS model as a particular example, the inverse
scattering transform we present in this paper for half-line problems can be readily generalized
to a wide range of integrable PDEs. The scattering system for the interval problems also lays the groundwork for an analytic method to solving  initial-boundary value problem for NLS on an interval.

The paper is organized as follows. In Section \ref{sec:1}, we provide the integrable aspects of the NLS equation under the periodic boundary conditions, and collect basic notions and notations needed in the rest of the paper. In Section \ref{sec:3}, Sklyanin's formalism of double-row monodromy matrix is  briefly discussed. Then, we provide a hierarchy of integrable boundary conditions. The scattering system for the double-row monodromy matrix is also established.  Section \ref{sec:4} deals with the inverse scattering transform for NLS on the  half-line using the ingredients of the previous sections. The main result is  stated in Theorem \ref{th:11} which is the Riemann-Hilbert formulation of the half-line problem. Connection to the nonlinear method of reflection is also provided. Examples of multi-soliton solutions of  NLS on the half-line are presented in Section \ref{sec:5}.

\section{Monodromy matrix and periodic problems for NLS}
\label{sec:1}
We recall some well-established results for the focusing NLS equation \eqref{eq:fnls}, and collect basic notions and notations needed in the rest of the paper. First, we present the notion of {\em monodromy matrix}, and construct the associated scattering system for the NLS equation subject to the {\em periodic boundary conditions}.  Then, by extending the fundamental domain of the periodic problems to the whole axis, we formulate the scattering system characterizing initial value problems for NLS on the whole line. Details and proofs can be found, for instance, in the monographes \cite{faddeev, NMSP}. The content of this section serves as a foundation for our development of the inverse scattering transform for the NLS equation on the half-line. 

The NLS equation \eqref{eq:fnls} is the result of the compatibility condition between
\begin{equation}
  \label{eq:laxp}
U\, \phi=\phi_x\,, \quad
   V\, \phi=\phi_t\,,
\end{equation} 
where $U$ and $ V$,  commonly known as the {\em Lax pair} of the NLS equation, are $2\times 2$  matrix-valued functions depending on the NLS fields, and a spectral parameter $\lambda$. They are in the forms (we drop the $x,t$ and $\lambda$ dependence for conciseness unless there is ambiguity) 
\begin{equation}\label{eq:UVee}
  U = -i \lambda \sigma_3 +Q \,,\quad V = -2i\lambda^2\sigma_3+2\lambda \, Q -iQ_x\,\sigma_3-i Q^2\,\sigma_3\,,
\end{equation}
with
\begin{equation}
  \label{eq:laxp2}
   Q=\bma 0 & q \\ -q^* & 0 \ema\,,\quad    \sigma_3=\bma 1 & 0 \\ 0& -1 \ema\,. 
\end{equation}
We call $U$ and $V$ respectively the {\em  space-part} and {\em time-part} of the Lax pair. They are traceless matrices, and obey  the  involution relations
\begin{equation}\label{eq:inv1}
  U^*(\lambda^*) = \sigma_2\,U(\lambda)\, \sigma_2  \,,\quad V^*({\lambda}^*) = \sigma_2\,V(\lambda)\, \sigma_2\,,\quad \sigma_2=\bma 0 & -i \\ i& 0 \ema\,,
\end{equation}
where the superscript  ${}^*$ denotes the complex conjugate.

Having the Lax pair, the key idea of the inverse scattering transform is to recast an initial-boundary value problem for NLS into a scattering system in the inverse space of $\lambda$. Consider, for example, the NLS equation subject to the periodic boundary conditions
\begin{equation}
  \label{eq:qbcs}
q(x,t) = q(x+L,t)\,. 
\end{equation}
The space variable $x$ is defined on a circle of length $L$. Alternatively, we  could also think of $x$ belonging to the whole axis by fixing the fundamental domain  to be the interval $
0 < x\leq L$.   The associated scattering system could be formulated by means of the so-called {\em  transition matrix}.

Define the transition matrix  $T(x,y;\lambda)$   as (we drop the dependence on $t$) 
\begin{equation}
  \label{eq:transm}
T(x,y;\lambda) = \overset{\curvearrowleft}{\exp}  \int^{x}_{y} U(\xi;\lambda)\, d\xi, \quad T(x,y;\lambda)\vert_{x = y} = I\,,
\end{equation}
where  $\overset{\curvearrowleft}{\exp}$ denotes the path-ordered exponential, and $I$  the identity matrix. The transition matrix is the fundamental solution of the space-part of the Lax equations, \ie the first equation in \eqref{eq:laxp}. It consists of  a formal integration along a path at a fixed time. By definition, one has
\begin{equation}
  \frac{\partial}{\partial x}T(x,y;\lambda) = U(x;\lambda)T(x,y;\lambda)\,,\quad   \frac{\partial}{\partial y}T(x,y;\lambda) = -T(x,y;\lambda)U(y;\lambda)\,,
\end{equation}
and  $ \det T(x,y;\lambda)  =1$ due to the traceless property of $U$. Assuming that the NLS field is a smooth function in $x$, then the (periodic)  monodromy matrix  $T_L( \lambda)$  can be defined as the transition matrix with the path of  integration going along the fundamental domain (see Fig.~\ref{fig:sm})
\begin{equation}\label{eq:srmm}
  T_L(t;\lambda) =  \overset{\curvearrowleft}{\exp}  \int^{L}_{0} U(\xi,t;\lambda)\, d\xi \, . 
\end{equation}
The monodromy matrix $T_L(t;\lambda)$ encodes the initial conditions of a given NLS field. The periodicity naturally leads to a Lax formulation  for $  T_L(t;\lambda)$. 
\begin{figure}[h]
  \centering
  \begin{tikzpicture}[scale=0.6]
    \path [draw=blue, postaction={on each segment={mid arrow=black}}]
    (0,0) ellipse(4 and 1.5);
    \pgfmathsetmacro \ll {2}
    \tikzstyle{nod}= [circle, inner sep=0pt, fill=black, minimum size=2pt, draw]
    \tikzstyle{nodb}= [circle, inner sep=0pt, fill=blue, minimum size=2pt, draw] 
    \draw[black] (0,0) ellipse (3 and 1.2);
    \node[nodb] (x0) at (-4,0) [label=below left:$0$] {};
    \node[nodb] (x0) at (-4,0) [label=above left:$L$] {};
    \node[nodb] (x0) at (-4,0) [label=left:$~$] {};
    \node[below] (x0) at (3.2,-1.2) {$T_L$};
  \end{tikzpicture}
  \caption{The monodromy matrix along the fundamental domain $0<x\leq L$}\label{fig:sm}\end{figure}
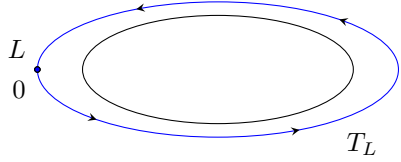
\begin{lemma}\label{lem:1}
For the NLS field satisfying the periodic boundary conditions \eqref{eq:qbcs}, one has 
\begin{equation}\label{eq:TVT}
\frac{d}{dt}T_L(t;\lambda) =[ V(L,t;\lambda),    T_L(t;\lambda))]\,,
\end{equation}which forms a  Lax pair. 
\end{lemma}
This Lax formulation is obtained by differentiating the monodromy matrix with respect to $t$
\begin{equation}
  \label{eq:41}
\frac{d}{dt}T_L(t;\lambda) = V(L,t;\lambda) T_L(t;\lambda) -  T_L(t;\lambda) V(0,t;\lambda)\,. 
\end{equation}
Then, the periodic conditions \eqref{eq:qbcs} yield \eqref{eq:TVT}. 
The Lax formulation \eqref{eq:TVT} implies that the quantity $\tr T_L(t;\lambda) $ could be considered as a generating function for infinitely many conserved quantities by taking $\tr T_L(t;\lambda) $ as a series expansion in $\lambda$. This indicates the integrability of the model.
\begin{rmk}

  Note that a similar Lax formulation as \eqref{eq:TVT} could be derived for the NLS equation subject to the quasi-periodic boundary conditions
  \begin{equation}
    \label{eq:2ql}
    q(x,t) = q(x+L,t)e^{i\varphi}\,,\quad 0\leq \varphi <2\pi\,,
  \end{equation}
 which represent a generalization of \eqref{eq:qbcs}. Based on the $r$-matrix formalism,  one could treat the periodic problems \eqref{eq:qbcs}, or more generally, the quasi-periodic problems  \eqref{eq:2ql},  for NLS as   integrable Hamiltonian field theories  \cite{faddeev}.
\end{rmk}
We briefly discuss the scattering system connected by the monodromy matrix $T_L(t;\lambda)$.
Let  $F_L(x;\lambda)$ and $G_L(x;\lambda)$ be two fundamental solutions of the space-part of the Lax equations  \eqref{eq:laxp}  in the form
  \begin{equation}\label{eq:qFG}
        F_L(x;\lambda) =  T(x,L;\lambda) \,,\quad         G_L(x;\lambda) =  T(x,0;\lambda) \,. 
  \end{equation}
  The function $G_L(x;\lambda)$ can be understood as $F_L(x;\lambda)$ acted by a shift operator shifting $F_L(x;\lambda)$ by $L$, since, by definition, one has
  \begin{equation}
    \label{eq:scatt6}
  G_L(x;\lambda) =  T(x,0;\lambda) =T(x+L,L;\lambda) =  F_L(x+L;\lambda)\,. 
  \end{equation}
It also follows from the definition of transition matrix that  $F_L$ and $G_L$ are connected by the monodromy matrix $T_L(\lambda)$ as
\begin{equation}
  \label{eq:s12}
  G_L(x;\lambda) =F_L(x;\lambda) T_L(\lambda)\,. 
\end{equation}
Note that the particular choice of the starting point of the scattering system \eqref{eq:s12} (in our case the point $x=0$) is irrelevant to the spectral property of $T_L(\lambda)$, and  could be brought to any  point in the fundamental domain \cite{N1}.  Once the scattering system is known, one can refer to the  elegant integration technique, konwn as finite-gap  method, which determines possible analytic and spectral properties of the scattering system \eqref{eq:s12}, and gives general expressions of the fundamental solutions as the (vector) Baker-Akhiezer functions for given spectral data \cite{N1, NMSP, DUR, IMs, AM1, bel1}. Inversely, this leads to analytic solutions of the periodic problems for NLS. 

Similarly, the scattering system for NLS on the whole axis could be formulated by extending the fundamental domain to the whole axis. Assuming the NLS field decays rapidly as $x\to \pm \infty$, which corresponds to the vanishing boundary conditions. Then, one can define two  fundamental solutions $F$ and $G$ as \begin{equation}\label{exsss}
  F(x;\lambda) =\lim_{y\to \infty}T(x,y;\lambda)e^{-i\lambda y\sigma_3}\,,\quad G(x;\lambda) =\lim_{y\to- \infty}T(x,y;\lambda)e^{-i\lambda y\sigma_3}\,,    \end{equation}
which are commonly known as {\em Jost solutions}  as ``extended'' versions of $F_L$ and $G_L$ defined in \eqref{eq:qFG}  by letting  $0 \to -\infty$ and $L \to\infty$.  The exponential terms appearing in \eqref{exsss} allow $F, G$ to have the well-defined asymptotic behaviors
\begin{equation}
  \label{eq:1fgfl}
\lim_{x\to \infty}  F(x;\lambda) = e^{-i\lambda x\sigma_3}  \,,\quad \lim_{x\to- \infty}  G(x;\lambda) = e^{-i\lambda x\sigma_3}  \,. 
\end{equation}
They are connected by the extended (full line)  monodromy matrix $T(\lambda)$ as 
\begin{equation}\label{flT} 
    G(x;\lambda) =F(x;\lambda) T(\lambda)\,,\quad T(\lambda)=  \lim_{y\to\infty} \left(e^{i\lambda y\sigma_3}\,T(y,-y;\lambda)  \,e^{i\lambda y\sigma_3} \right)\, . 
\end{equation}
Therefore, the NLS equation on the full line could be interpreted as a periodic problem by gluing the two infinity points $\pm \infty$ together.  The inverse scattering transform is based on analysis of the scattering system, which eventually leads to solutions of an initial value problem for NLS. In particular, multi-soliton solutions can be constructed as special solutions of the model.

\section{NLS on an  interval: Sklyanin's formalism and hierarchy of integrable boundary conditions}\label{sec:3}
In this section, we consider the focusing NLS equation \eqref{eq:fnls} restricted on a closed interval. Based on Sklyanin's formalism of double-row monodromy matrix, we provide a set of sufficient criteria (see Lemma~\ref{lem:kkpm}) leading to a Lax formulation for the double-row monodromy matrix. These criteria correspond to constraints involving the so-called reflection matrix and the time-part of the Lax pair of NLS. This gives rise to a hierarchy of reflection matrices, which is accompanied with a hierarchy of integrable boundary conditions. Explicit examples of reflection matrices and the associated boundary conditions in the hierarchy are derived, including some new ``implicit  boundary conditions'' for NLS. The scattering system for the double-row monodromy matrix is also formulated, which characterizes the interval problem for NLS in the inverse space. Comparisons between the interval problems  and various existing methods are also provided.
\subsection{Double-row monodromy matrix}
\begin{figure}[h]\centering  \begin{tikzpicture}[scale=0.6]
    \tikzstyle{nod}= [circle, inner sep=0pt, fill=black, minimum size=2pt, draw]
    \tikzstyle{nodb}= [circle, inner sep=0pt, fill=blue, minimum size=2pt, draw]
    \draw[thick, black] (-2.5,0) -- (2.5,0);
    \draw[thick, black] (-2.5,0.15) -- (-2.5,-0.15);
    \draw[thick, black] (2.5,0.15) -- (2.5,-0.15);
    \node[left] (x0) at (-2.5,0)  {$0$};
    \node[right] (x0) at (2.5,0)  {$L$};
    \node[nodb] (x0) at  (3.41,0) [label=right:$K_+$] {};
    \node[left] (x0) at  (-3.41,0) {$K_-$};
    \node[below](a) at (0,-0.7){$ T_L $};
    \node[above](a) at (0,0.7){$ T_L^{-1} $};
  \path [draw=blue,postaction={on each segment={mid arrow=black}}]
  (-3,-0.7) to  (3,-0.7)
  (3,0.7) to   (-3,0.7)
  (3,-0.7)to[out=0,in=0]  (3,0.7)
  (-3,0.7)to[out=-180,in=-180]  (-3,-0.7)
  ;

  \end{tikzpicture}
  \caption{The double-row monodromy matrix following the path circling the interval.}
  \label{fig:monomatrix}
\end{figure}
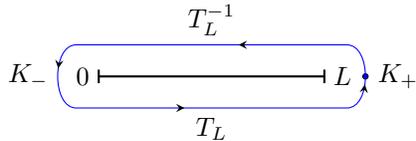

Let the NLS equation \eqref{eq:fnls} be defined on an interval $0< x< L$ subject to certain boundary conditions that will be determined at the two endpoints $x=0$ and $x=L$. The key idea of Sklyanin's approach  to dealing with this kind of initial-boundary value problems, and to selecting admissible class of boundary conditions which preserve the integrability of the model, is to formulate a monodromy matrix integrated following a path circling the interval by employing both the ``single-row'' transition matrix $T_L$ and the reflection matrices $K_\pm$ \cite{SKBC, sklyanin1988boundary}. This construction is illustrated in Fig.~\ref{fig:monomatrix}. We call this type of monodromy matrix  {\em double-row monodromy matrix}, {\it cf}.~\cite{ACC}. The boundary conditions of NLS are encoded into certain constraints involving the time part of the Lax pair of NLS and reflection matrices $K_\pm$.

Let the double-row monodromy matrix $\Gamma_L(t;\lambda)$ be defined as
\begin{equation}\label{eq:drmm1}
  \Gamma_L(t;\lambda) = T_L(t;\lambda)K_-(t;-\lambda)T^{-1}_L(t;-\lambda)K_+(t;\lambda)\,,
\end{equation}
where $T_L$ is the single-row transition matrix defined in \eqref{eq:srmm}, and the $2\times 2$ matrices $K_+$ and $K_-$ are  the reflection matrices associated to the boundaries $x=L$ and $x=0$ respectively. The appearance of $-\lambda$ in $\Gamma_L(t;\lambda)$ is   a characteristic feature to the NLS model \cite{SKBC}, and is of crucial importance in the formulation of scattering system. This can be interpreted as a discrete action of  the reflection matrices:
 the action of $K_\pm$ on an auxiliary field is accompanied by a $\ZZ_2$ action transforming the spectral parameter $\lambda$ to $-\lambda$. 
 \begin{rmk}
   \label{rmk:dr}
   Note that the choice of the starting point of the path of integration for the double-row monodromy matrix,  as well as the orientation of the path, is irrelevant to the spectral properties of $\Gamma_L(t;\lambda)$. As to our definition \eqref{eq:drmm1}, we fix the starting point at $x=L$ and make the path anti-clockwise. This choice will be particularly convenient for the formulation of the scattering system for the NLS equation on the half-line (see Section~\ref{sec:4}).
 \end{rmk}
We will assume  the reflection matrices $K_\pm$ to be  non-degenerate, but time-dependent. The non-degeneracy of $K_\pm$ ensures that $\Gamma_L(t;\lambda)$ is also non-degenerate.  The following Lemma provide a set of sufficient conditions for the existence of a Lax formulation for  the double-row monodromy matrix. 
\begin{lemma}\label{lem:kkpm}Assume $K_+$ and $K_-$ to be non-degenerate  $2\times 2$ matrices. If they satisfy
  \begin{subequations}
\label{eq:rmpm}  \begin{align}
        \frac{d}{dt}K_+(t;\lambda)& = V(L,t;-\lambda)K_+(t;\lambda)-K_+(t;\lambda)V(L,t;\lambda)\,, \label{eq:k2}\\
    \frac{d}{dt}K_{-}(t;\lambda)& = V(0,t;-\lambda)K_-(t;\lambda)-K_-(t;\lambda)V(0,t;\lambda)\,, \label{eq:k1}
  \end{align}    
  \end{subequations}
where $V$ is the time-part Lax matrix of NLS defined in \eqref{eq:UVee},  then
  \begin{equation}\label{eq:gammat}
\frac{d}{dt}\,\Gamma_L(t;\lambda) = [V(L,t;\lambda),\Gamma_L(t;\lambda)]\,,
\end{equation}
which forms a  Lax pair. 
\end{lemma}
This result can be proved using straightforward computations. Clearly,  the quantity $\text{tr}\,\Gamma_L(t;\lambda)$ is preserved by
the time evolution, and can be interpreted as a generating function for infinitely many conserved quantities following a series expansion in $\lambda$. A  Hamiltonian picture  of the Lax pair \eqref{eq:gammat} based on the $r$-matrix approach was recently established  in \cite{ACC}.  

In Lemma~\ref{lem:kkpm}, the reflection matrices $K_+$ and $K_-$ obey exactly the same type of equations, and are independent to each other, since the two equations in \eqref{eq:rmpm} are defined at places. Before tackling these equations in details, we characterize some essential properties of $K_\pm$. 
\begin{description}
\item[{\bf \em  1) Semi-discrete Lax pair interpretation.}]  Consider the gauge-type transformation \eqref{eq:rmpm} (we drop the subscripts and assume the space variable $x$ being one of the endpoints of the interval)
  \begin{equation}\label{eq:kvk}
    \frac{d}{dt}K(t;\lambda) = V(t;-\lambda)K(t;\lambda) -K(t;\lambda) V(t;\lambda)\,.
  \end{equation}
This equation can be seen as a semi-discrete Lax pair, with the understanding that the action of the reflection matrix $K$ is accompanied by a discrete action transforming $\lambda$ to $-\lambda$. Let
  \begin{equation}\label{eq:KETA}
    K(t;\lambda)\,\phi(t;\lambda) = \eta\,  \phi(t;-\lambda)\,,\quad \frac{d}{dt}\phi(t;\lambda)=V(t;\lambda)\,\phi(t;\lambda)\,, 
  \end{equation}
  where $\eta$ is some time-independent parameter\footnote{The validity of this form is argued in Remark~\ref{rmk:31}.}, then the compatibility yields \eqref{eq:kvk}. Now the question is: having $V$ that is the time-part Lax matrix of NLS, what are the possible forms of $K$ subject to \eqref{eq:kvk}, and what are the associated boundary conditions for NLS?
\item[{\bf \em  2) The determinants of  $K_\pm$ are time-independent.}] This is a direct consequence of  the Lax formulation \eqref{eq:gammat}. 
 By integrating  \eqref{eq:gammat}, one has the general expression of $\Gamma_L(t;\lambda)$ as \begin{equation}
  \Gamma_L(t;\lambda) =S(t,t_0;\lambda)\Gamma_L(t_0;\lambda)S(t_0,t;\lambda)\,,\quad S(t_2,t_1;\lambda): =    \overset{\curvearrowleft}{\exp}  \int^{t_2}_{t_1} V(L,\tau,\lambda)\, d\tau \, , 
\end{equation}
which implies  that the determinant of $ \Gamma_L(t;\lambda)$ is time-independent. Since 
\begin{equation}
  \label{eq:17}
  \det \Gamma_L = \det K_- \det K_+\,,
\end{equation}
 and $\det K_+$ and $\det K_-$ are independent to each other, they are time-independent.

\item[{\bf \em  3) Normalization property  (R).}] We use  $\kappa_\pm(\lambda)$ to denote $\det K_\pm(t;\lambda)$. Moreover, we require that $K_\pm$ are normalized as\footnote{The first relation in \eqref{eq:ks} can be derived using  \[\frac{d}{dt}K^{-1}(t;\lambda) =V(t;\lambda)K^{-1}(t;\lambda) -K^{-1}(t;\lambda)V(t;-\lambda)\,,\] which implies  $K^{-1}(t;\lambda)$ and  $K(t;-\lambda)$ obey the same constraint \eqref{eq:kvk}.  The second relation is a consequence of the involution property \eqref{eq:inv1} of the Lax matrix $V$. 
}
\begin{equation}
\label{eq:ks}    K_{\pm}^{-1}(t;\lambda) = K_{\pm}(t;-\lambda)\,,\quad \kappa_{\pm}(-\lambda)K_{\pm}(t;\lambda)=\sigma_2{K}_{\pm}^*(t;{\lambda}^*)\sigma_2\,,
  \end{equation}  
which is equivalent to fix $\kappa_\pm(\lambda)$ as    \begin{equation}\label{eq:ff1}
    \kappa_\pm(\lambda)\kappa_\pm(-\lambda) =1\,,\quad {\kappa}_\pm^*({\lambda}^*) =\kappa_\pm(-\lambda)\,.
  \end{equation}
This is called normalization property (R), and will be extensively used in the rest of the  paper. 
\end{description}


\subsection{Hierarchy of reflection matrices and integrable boundary conditions}
\label{sec.32}  We explore the constraint \eqref{eq:kvk}, and derive possible forms to the reflection matrix $K$ and the associated boundary conditions. 
We follow the normalization property (R), and assume that $\kappa(\lambda)$ is a rational function of $\lambda$. Then,  the normalization \eqref{eq:ff1} gives rise to the following representation of $\kappa(\lambda)$ 
\begin{equation}\label{eq:normk1}
\kappa(\lambda) = \epsilon \frac{f(\lambda)}{g(\lambda)}= \epsilon \prod_{j=1}^{\cN}\left[\frac{\lambda-\beta_{j}}{\lambda-\beta_{j}^{*}}\frac{\lambda+\beta_{j}^{*}}{\lambda+\beta_{j}}\right]\prod_{\ell=1}^{\cM}\left[\frac{\lambda-i\alpha_\ell}{\lambda+i\alpha_\ell}\right]\,,\quad \epsilon =\pm 1\,,
\end{equation}
where
\begin{equation}
  \label{eq:14}
 f(\lambda) = \prod_{j=1}^{\cN}\left[(\lambda-\beta_{j})(\lambda+\beta^*_{j})\right]\prod_{\ell=1}^{\cM}\left[\lambda-i\alpha_\ell\right]\,,\quad g(\lambda) =  f^*(\lambda^*)\,,
\end{equation}
for given positive integers $\cN$ and $\cM$.
It is required that the parameters $\beta_j$, $j=1,\dots, \cN$, are complex numbers with nonzero real and imaginary parts, {\it ie}.~$\text{Re}\, \beta_j\neq 0$ and $\text{Im}\, \beta_j\neq 0$; and  $\alpha_\ell$, $\ell = 1,\dots, \cM$,  are nonzero real numbers,  {\it ie}.~$\alpha_\ell \neq 0$. We take \eqref{eq:normk1} as a definition of $\kappa(\lambda)$. There is an extra (parity) parameter $\epsilon = \pm 1$ appearing in  \eqref{eq:normk1}, and clearly
\begin{equation}
  \lim_{\lambda\to \infty} \kappa(\lambda)   = \epsilon\,.
\end{equation}

We use  $\fD$ to denote  the set of zeros of $\kappa(\lambda)$ (hence zeros of $f(\lambda)$)
\begin{equation}\label{eq:fd1}
{\fD } = \{  \lambda \in \CC ~\text{such that}~\kappa(\lambda) =0\} = \{\beta_j, -\beta^*_j\}_{j=1,\dots, N}\cup \{i\alpha_\ell\}_{\ell= 1, \dots, M}\,. 
\end{equation}
Similarly,  $\fD^*$ denotes the set of poles of $\kappa(\lambda)$ (hence zeros of $g(\lambda)$), whose elements are the complex conjugate of those of $\fD $.   Therefore,  $\kappa(\lambda)$ is completely determined by a given set $\fD$ as well as a parity parameter $\epsilon = \pm 1$. In general,  there is no restriction on the multiplicity of zeros (and poles) for  $\kappa(\lambda)$. 

Given $\kappa(\lambda)$  in the form \eqref{eq:normk1}, we are looking for possible forms of $K$ subject to the constraint \eqref{eq:kvk}.
Introduce another matrix-valued function $\cL$ in the form
\begin{equation}
\label{eq:LKa}\cL(\lambda) = g(\lambda)K(\lambda) = \begin{pmatrix} \cA(\lambda)&\cB(\lambda)\\\cC(\lambda)&\cD(\lambda)\end{pmatrix} \,,
\end{equation}
where $g(\lambda)$ is the denominator of $\kappa(\lambda)$ as defined in \eqref{eq:normk1}. 
Clearly, $\cL$ also satisfies \eqref{eq:kvk}, and its entries are polynomial of $\lambda$ of degree $\fN=2\cN+\cM$. Write down   $\cL(\lambda)$  in the form
\begin{equation}
  \label{eq:10cl}
  \cL(\lambda)  =\lambda^{\fN}   \cL_0+\lambda^{\fN-1}   \cL_1+\cdots+   \lambda \cL_{\fN-1}+    \cL_{\fN}\,,\quad  \cL_{j} =   \begin{pmatrix} \cA_j&\cB_j\\\cC_j&\cD_j\end{pmatrix}\,,\quad j =1 ,\dots, \fN\,.
\end{equation}
The relations between the entries of $\cL$ are presented in Table~\ref{table:abcd}. 
Plugging this form of $\cL$ into the constraint \eqref{eq:kvk} yields a hierarchy of reflection matrices. This also leads to a hierarchy of boundary conditions.
\begin{table}[h!]
\begin{center}
\renewcommand\arraystretch{1.5}
\begin{tabular}{|c|c|c|c|}
\hline
 & odd $\fN$ (odd $\cM$)   & even $\fN$ (even $\cM$) \\
\hline
$\epsilon=\pm 1$ & \begin{tabular}{c} $\cD(\lambda)=-\epsilon\, \cA(-\lambda)\,,~\cD(\lambda)= \epsilon\, \cA^*(\lambda^*)$ \\ $\cB(-\lambda)=\epsilon\,\cB(\lambda)\,,~\cC(\lambda)=-\epsilon\,\cB^*(\lambda^*)$ \end{tabular} & \begin{tabular}{c} $\cD(\lambda)=\epsilon\,\cA(-\lambda)\,,~\cD(\lambda)=\epsilon\, \cA^*(\lambda^*)$ \\ $\cB(-\lambda)=-\epsilon\,\cB(\lambda)\,,~\cC(\lambda)=-\epsilon\,\cB^*(\lambda^*)$ \end{tabular} \\
 \hline
\end{tabular}
\caption{Relations between the entries of $\cL$ due to the normalization property (R). }
\label{table:abcd}
\end{center}
\end{table}

There is a combinatorial aspect in determining the degree $\fN$ as well as the expressions of $\cL$ (and $K$), since $\fN$ depends on  $\cN$ and $\cM$ for $\fN\geq 2$, and $\kappa(\lambda)$ depends on an extra parity parameter $\epsilon$.
There is also a sign freedom for $K$:  given $K $ as a solution of \eqref{eq:kvk},  $-K$ is also a solution with the same the normalization.  We  fix this sign freedom by requiring (since the leading term  $\cL_0$ in \eqref{eq:10cl} must be a diagonal constant matrix)
\begin{equation}\label{eq:asymK}
  \lim_{\lambda \to \infty} K(\lambda) = \bma 1 &0 \\ 0& \epsilon \ema\,.
\end{equation}
This form will be needed in Section~\ref{sec:4} when dealing with the uniqueness of solutions of the Riemann-Hilbert problems.

\begin{rmk}\label{rmk:31}
It is not entirely new in the literature to consider the gauge-type constraint \eqref{eq:kvk} with a time-dependent reflection matrix to derive integrable boundary conditions (or integrable defect conditions in some context), {\it cf}. \cite{ BM1, cz11,Causyst, ZAMBON}. However, to the best of the authors' knowledge,  the point of view that the constraint \eqref{eq:kvk} is treated as a semi-discrete Lax pair \eqref{eq:KETA} possessing a hierarchy of reflection matrices accompanied with a hierarchy of integrable boundary conditions has not been fully considered.  
\end{rmk}

In the following, we show some explicit examples for $\fN =1, 2, 3$. Except for one particular case (Case $1$), one has, in general, time-dependent reflection matrices as solutions of \eqref{eq:kvk}, which are accompanied by time-dependent boundary conditions.   
Since we are dealing with integrable models,  the NLS fields are assumed to be smooth enough to be compatible with the bulk equation when the space variable is approaching to  the boundary point. 
The computations are becoming increasingly complicated as  $\fN$ increases. For $\fN\geq 2$, it turns out that in some cases the coefficient functions  $\cA_j,\cB_j,\cC_j,\cD_j$  are involved in certain nonlinear differential-algebraic constraints, which do not admit explicit expressions in terms of the NLS fields. This kind of constraints will be referred to as {\em implicit boundary conditions}. They can be understood as coupled nonlinear differential-algebraic systems with some additional fields at the boundary to be determined. 
The question of whether there exists a systematic approach to deriving higher-order reflection matrices and the associated boundary conditions based on some recursion-type relations  remains open. 
\begin{description}

\item[Case $1$, $\fN =1, \cN = 0, \cM =1, \epsilon= -1$, Robin boundary conditions.] The reflection matrix and the boundary conditions can be  fixed as \begin{equation}
\label{eq:Rb1}K(\lambda)=\frac{1}{\lambda+i\alpha_1}\begin{pmatrix}
\lambda+\cA_1 & 0 \\ 0 & \cA_1- \lambda
\end{pmatrix} \,, \quad q_x =  2 i\cA_1 q\,, 
\end{equation}where $\cA_1$ is  a pure imaginary number satisfying $\cA_1^2 = -\alpha_1^2$.  This case was originally derived by Sklyanin in \cite{SKBC}, and reappeared in various contexts, {\it cf}.~\cite{Tarasov, HH1, fokas2002integrable}. The (zero) Dirichlet and (zero) Neumann boundary conditions (we assume that the boundary point is located at $x=0$), \ie
\begin{equation}
  q(0,t) =0 \,,\quad q_x(x,t)\vert_{x=0} = 0\,, 
\end{equation}
can be respectively obtained as limiting cases as $\alpha_1 \to \infty$ and $\alpha_1 \to 0$.
\item[Case $2$, $\fN =1, \cN = 0, \cM =1, \epsilon= 1$, ``defect-type'' boundary conditions.] We obtain 
\begin{equation}\label{eq:case21}
K(\lambda)=\dfrac{1}{\lambda+i\alpha_1}\begin{pmatrix}
\lambda +i\Omega & iq \\iq^* & \lambda- i\Omega
\end{pmatrix}\,,\quad    iq_t+2|q|^2q- 2q_x\Omega =0 \,, 
\end{equation}
where  $\Omega^2=\alpha_1^{2}-|q|^2$ with  $|q| \leq |\alpha_1| $. The boundary conditions can be associated with the integrable defect conditions for NLS derived in  \cite{Causyst}\footnote{In \cite{Causyst}, the ``new'' integrable defect conditions at a fixed point (assumed to be $x=0$) for NLS read 
  \[
    (q-\widetilde{q})_x   = i \beta (q-\widetilde{q}) - \Omega(q+\widetilde{q})\,,\quad
    (q-\widetilde{q})_t   = - \beta (q-\widetilde{q})_x -i \Omega(q+\widetilde{q})_x + i(q-\widetilde{q})(|q|^2+|\widetilde{q}|^2)\,, 
\]  where 
  \[
  \Omega^2 = \alpha^{2}-\frac{1}{2}(|q|^2+|\widetilde{q}|^2)\,,   \]
and $\alpha, \beta$ are real parameters. The NLS fields $q$ and $\widetilde{q} $, connected by the above defect conditions, are living respectively in  the positive and negative semi-axis. Letting $\beta = 0$ and $\widetilde{q}(x) = -q(-x)$, one recovers  \eqref{eq:case21}.}. 
The limiting cases $\alpha_1 \to 0$ and $\alpha_1 \to \infty$ correspond respectively to the Dirichlet and Neumann boundary conditions. 

\item[Case $3$, $\fN =2, \cN = 1, \cM =0, \epsilon= 1$, ``new'' integrable boundary conditions  \cite{ZAMBON}.] 
   \begin{equation}
K(\lambda)= \dfrac{1}{(\lambda+\beta_1)(\lambda-\beta_1^{*})} \begin{pmatrix}
  \lambda^{2}+ i\lambda \Omega+\Pi & iq\lambda \\iq^*\lambda & \lambda^{2}-i\lambda \Omega+\Pi
\end{pmatrix} \,,
\end{equation}
where $\Omega^2=-(\beta_1-\beta_1^{*})^{2}-|q|^2$ with $\Omega$  being real and $\Pi^2 = |\beta_1|^4$. The boundary conditions read
\begin{equation}
\label{eq:iqt2}iq_t+2|q|^2q- 2q_x\Omega+4q\Pi =0 \,,
\end{equation}
with $ |q| \leq 2| \text{Im}\,\beta_1| $. The boundary conditions \eqref{eq:iqt2} were first obtained in \cite{ZAMBON}  by dressing a Dirichlet boundary condition by some integrable defect conditions. Soliton solutions of half-line problems for NLS equipped with these boundary conditions were recently derived in \cite{Gruner2, Xia}.  
\item[Case $4$,  $\fN =2, \cN = 0, \cM =2, \epsilon= 1$, deformed versions of \eqref{eq:iqt2}.] One has
\begin{equation}
K(\lambda)= \dfrac{1}{(\lambda+i\alpha_1)(\lambda+i\alpha_2)} \begin{pmatrix}
\lambda^{2}+ i\lambda \Omega+\Pi & iq\lambda \\iq^*\lambda & \lambda^{2}- i\lambda \Omega+\Pi
\end{pmatrix} \,,
\end{equation}where $\Omega^2=(\alpha_1+\alpha_2)^{2}-|q|^2$ with $\Omega$  being real and $\Pi^2 = \alpha_1^2\alpha_2^2$. The boundary conditions read \begin{equation}
\label{eq:eq:iqt2}iq_t+2|q|^2q- 2q_x\Omega+4q \Pi =0 \,,
\end{equation}
with $ |q| \leq | \alpha_1+\alpha_2| $. Note that \eqref{eq:eq:iqt2} and \eqref{eq:iqt2} are in similar forms. However, they are associated with different sets of parameters, and cannot be mapped between each other. 
\item [Case 5, $\fN =2, \cN = 1, \cM =0, \epsilon= -1$, implicit boundary conditions I.] One has
\begin{equation}
K(\lambda) = \dfrac{1}{(\lambda+\beta_1)(\lambda-\beta_1^{*})} \begin{pmatrix} \lambda^2+\cA_1\lambda+\cA_2 & \cB_2 \\ \cB_2^* & -\lambda^2+\cA_1\lambda-\cA_2 \end{pmatrix} \,,
\end{equation}
with $\cA_1^{*}=-\cA_1$, $ \cA_2^{*}=\cA_2$. The boundary conditions are implicitly defined as 
\begin{equation} 2q\cA_1+2i\cB_2+iq_x=0\,,
\end{equation}
where  $\cA_1,\cA_2, \cB_2$ are subject to the following nonlinear algebraic-differential constraints
\begin{equation}
\cA_1^2-2\cA_2=\beta_1^2+(\beta_1^*)^2\,,\quad
\cA_2^2+|\cB_2|^2=|\beta_1|^4\,,\quad \frac{d}{dt}\cB_2=2i|q|^2\cB_2-2iq_x\cA_2\,.  
\end{equation}
The last equation (and its  complex conjugate) is a nonlinear ordinary differential equation for $\cB_2$ by taking $\cA_1, \cA_2$ as expressions of $\cB_2,\cB_2^*$ and parameters (using the first two equations). The quantities $\cB_2$ and $\cB_2^* $ can be understood as ``additional fields'' present at the boundary. 

\item[Case 6,  $\fN =3, \cN = 1, \cM =1, \epsilon= 1$, implicit boundary conditions II.]
One has
\begin{equation}
K(\lambda) \propto \begin{pmatrix} \lambda^3+\cA_1\lambda^2+\cA_2\lambda+\cA_3 & i q\lambda^2+\cB_3 \\ iq^{*}\lambda^2-\cB_3^* & \lambda^3-\cA_1\lambda^2+\cA_2\lambda-\cA_3 \end{pmatrix} \,,
\end{equation}
where $\cA_1^{*}=-\cA_1$, $\cA_2^{*}=\cA_2$, $\cA_3^{*}=-\cA_3$, and   $\cA_1,\cA_2, \cA_3, \cB_3$   satisfy
\begin{subequations}
\label{det}
\begin{align}
\cA_1^2-2\cA_2-|q|^2&=\beta_1^2+(\beta_1^*)^2-\alpha_1^2 \,,
\\
\cA_2^2-2\cA_1\cA_3+i(q\cB_3^*-q^*\cB_3)&=-\alpha_1^2\beta_1^2-\alpha_1^2(\beta_1^*)^2+|\beta_1|^4 \,,
\\
\label{a3b3}
\cA_3^2-|\cB_3|^2&=-\alpha_1^2|\beta_1|^4 \,,\\\dfrac{d}{dt}\cB_3&=2i(|q|^2\cB_3-q_x\cA_3) \,.
\end{align}
\end{subequations}
Again the term $\cB_3$ here is an additional field present at the boundary. 
Then, the boundary conditions are implicitly defined as
  \begin{equation}
iq_t+4q\cA_2+2|q|^2q+4i\cB_3+2iq_x\cA_1 = 0 \,. 
\end{equation}
\end{description}

One could obtain another case for  $\fN =2$ with $\cN = 0, \cM =2, \epsilon= -1$  which corresponds to a deformed version of Case $5$. For the half-line NLS model considered in this paper, based on the inverse scattering transform developed in Section \ref{sec:4}, the reflection matrix $K$ can be reconstructed as  a dressed form of certain constant reflection matrix (see Corollary~\ref{cor:11} in Section~\ref{sec:43}).  This means that  the additional fields, for instance, $\cB_2$ in Case $5$ appearing in the implicit boundary conditions I, can be explicitly computed for the half-line problem we consider. 

\begin{rmk}
 Let us comment on the validity of the semi-discrete Lax \eqref{eq:KETA}.  We follow the possible forms of $K$ discussed above. Set the parameter $\eta$  in \eqref{eq:KETA} to be $\kappa(\lambda)$ (we do not touch the time-deformation part, hence drop the time dependence), and let $\phi(\lambda)$ and $\phi(-\lambda)$ be in the forms
  \begin{equation}
    \phi(\lambda) = \bma  v(\lambda) \\ 1\ema \,,\quad     \phi(-\lambda) = \bma u(\lambda) \\ 1\ema\,,
  \end{equation}
One needs to prove  $u(\lambda) = v(-\lambda)$. Take $\kappa(\lambda)$ in the form $\kappa(\lambda) =  \epsilon f(\lambda)/g(\lambda)$ as defined in \eqref{eq:normk1}.  If $\cB$ and $\cC$ are nonzero quantities, then it follows from  \eqref{eq:LKa} that
  \begin{equation}
    v(\lambda) =  \frac{\epsilon f(\lambda) -\cD(\lambda)}{\cC(\lambda)}\,,\quad     u(\lambda) =  \frac{\cA(\lambda)-g(\lambda)}{\cC(\lambda)}\,. 
  \end{equation}
Moreover, one has $f(\lambda) =  (- 1)^\cM g(-\lambda)$, $\cD(\lambda) =  (- 1)^\cM\epsilon \cA(-\lambda)$ and $\cC(\lambda) = - (- 1)^\cM\epsilon\, \cC(-\lambda)$,  which amounts to the desired result. In  Case $1$,  $\cB$ and $\cC$ are zero  quantities,  one can choose for instance $v(\lambda) =i\alpha_1-\lambda$ and $u(\lambda) =i\alpha_1+\lambda$.  
\end{rmk}

\subsection{Scattering system  for the double-row monodromy matrix}
\label{sec.33}
Let us formulate the scattering system for the double-row monodromy matrix. 
This  lays the groundwork for direct scattering transform for the interval problems for the  NLS equation. We assume that  $K_\pm$ belong to the hierarchy of the reflection matrices explained previously.  The boundary conditions  at the  endpoints $x=L$ and $x=0$ are   respectively determined by the constraints \eqref{eq:k2} and \eqref{eq:k1}.  

We follow the path of double-row monodromy matrix as shown in Fig.~\ref{fig:monomatrix}.  Let $x$ be a particular point on the interval cutting the path of the double-row monodromy matrix as illustrated in Fig.~\ref{fig:monomatrix1}. 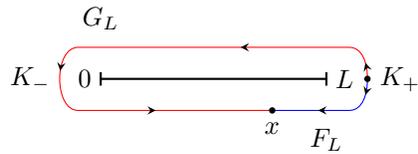
\begin{figure}[h]
  \centering
  \begin{tikzpicture}[scale=0.6]
    \tikzstyle{nod}= [circle, inner sep=0pt, fill=black, minimum size=2pt, draw]
    \tikzstyle{nodb}= [circle, inner sep=0pt, fill=blue, minimum size=2pt, draw]
    \draw[thick, black] (-2.5,0) -- (2.5,0);
    \draw[thick, black] (-2.5,0.15) -- (-2.5,-0.15);
    \draw[thick, black] (2.5,0.15) -- (2.5,-0.15);

    \node[left] (x0) at (-2.5,0)  {$0$};
    \node[right] (x0) at (2.5,0)  {$L$};
    \node[left] (x0) at  (-3.41,0) {$K_-$};

  \path [draw=blue,postaction={on each segment={mid arrow=black}}]
  (3.41,0)  to [out=-90,in=0]  (3,-0.7)
   (3,-0.7) to (1.3,-0.7)
  ;
  \path [draw=red,postaction={on each segment={mid arrow=black}}]
  (3.41,0)  to [out=90,in=0]  (3,0.7)
  (3,0.7) to (-3,0.7)
  (-3,0.7)to[out=-180,in=-180]  (-3,-0.7)
  (-3,-0.7) to  (1.3,-0.7)
  ;
    \node[nod] (x0) at (3.41,0) [label=right:$K_+$] {};
    \node[nod] (x0) at (1.3,-0.7) [label=below:$ x $]{};
    \node at (2.5,-0.7) [label=below:$ F_L$]{};
    \node at (-2.5,2) [label=below:$ G_L$]{};
  \end{tikzpicture}
  \caption{A scattering system connected by the double-row monodromy matrix $\Gamma$. The two fundamental solutions  $F_L,G_L$ are integrated from $x=L$ along two different paths.  }
  \label{fig:monomatrix1}
\end{figure}We also assume that the NLS field is a smooth function at a fixed time (we drop the dependence on $t$ for convenience). Define  two fundamental solutions $F_L(x;\lambda)$ and $G_L(x;\lambda)$ of the space-part of the Lax equations \eqref{eq:laxp} as 
\begin{equation}\label{eq:FGL}
    F_L(x;\lambda) = T(x,L;\lambda)\,,\quad G_L(x;\lambda)= T(x,0;\lambda)K_-(-\lambda)T_L^{-1}(-\lambda)K_+(\lambda)\,,
  \end{equation}
  where $T(x,y;\lambda)$ is the transition matrix defined in  \eqref{eq:transm}, and $T_L(\lambda)$ is the single-row transition matrix defined in \eqref{eq:srmm}. Following Fig.~\ref{fig:monomatrix1},  $F_L$ and $G_L$ are connected by the double-row monodromy matrix $\Gamma_L(\lambda)$ \eqref{eq:drmm1} as  \begin{equation}\label{eq:drmm11}
G_L(x;\lambda) =   F_L(x;\lambda)\,\Gamma_L(\lambda)\,. 
\end{equation}   
Define an auxiliary single-row transition matrix  $H_L(\lambda)$ as
\begin{equation}
  \label{eq:13H}
H_L(\lambda) =  K_-(-\lambda)T_L^{-1}(-\lambda)K_+(\lambda)\,,
\end{equation}
then $G_L(x;\lambda) =T(x,0;\lambda)H_L(\lambda)$, and $F_L$ and $G_L$ obey the following well-defined  behaviors at the boundary points
\begin{subequations}
\begin{align}
  \label{eq:16}
F_L(0;\lambda) = T^{-1}_L(\lambda)\,,\quad  &F_L(L;\lambda) = I\,,  \\
  G_L(0;\lambda) = H_L(\lambda)\,,\quad & G_L(L;\lambda) = \Gamma_L(\lambda)\,.
\end{align}  
\end{subequations}


Analysis of the scattering system \eqref{eq:drmm11}  could determine the spectral properties of the double-row monodromy matrix, and fix the analytic properties of the fundamental solutions. Inversely, this would lead to analytic solutions of the interval problems for NLS. This tantalizing project is beyond the scope of this paper, since it involves a rather different set of mathematical tools than that of the half-line problems. As to  NLS on the half-line, our approach relies on extending one of the endpoints of the interval to infinity as in the case of full line problems by  extending the fundamental domain of the periodic problems to the whole axis.

Here, we would like to clarify a number of important aspects regarding the use of Sklyanin's double-row monodromy matrix and emphasize the significance of the scattering system \eqref{eq:drmm11}. 

\begin{itemize}
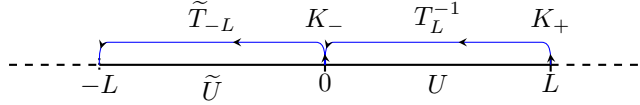
\begin{figure}[h!]\centering  \begin{tikzpicture}[scale=0.6]
    \tikzstyle{nod}= [circle, inner sep=0pt, fill=black, minimum size=2pt, draw]
    \tikzstyle{nodb}= [circle, inner sep=0pt, fill=blue, minimum size=2pt, draw]
    \draw[thick, black] (-7.5,0) 
    node[below] {$-L$} -- node[below] {$\widetilde{U}$}(-2.5,0);
    \draw[thick, black] (-2.5,0) -- node[below] {$U$}(2.5,0);
    \draw[thick, dashed, black] (2.5,0) -- (4.5,0);
    \draw[thick, dashed, black] (-9.5,0) -- (-7.5,0);
    \draw[thick, densely dotted, black] (-7.5,0.15) -- (-7.5,-0.15);
    \draw[thick, black] (2.5,0.15) -- (2.5,-0.15);
    \draw[thick, black] (-2.5,0.15) -- (-2.5,-0.15);
    \node[below] (x0) at (-2.5,0)  {$0$};
    \node[below] (x0) at (2.5,0)  {$L$};
    \node[above](a) at (2.5, .5){$ K_+ $};
    \node[above](a) at (-2.5, .5){$ K_- $};
    \path [draw=blue,postaction={on each segment={mid arrow=black}}]
(2.5, .0) to[out=90,in=0]   (2.3, .5) -- node[above] {$T_L^{-1}$}   (-2.3, .5)  to[out=180,in=90]  (-2.5, 0) to[out=90,in=0] (-2.7, .5)   --node[above] {$\widetilde{T}_{-L}$}  (-7.3, .5) to[out=180,in=90] (-7.5, 0);
  \end{tikzpicture}
  \caption{Periodic extension of the interval problem.}
  \label{fig:monomatrixsddd}
\end{figure}
\item {\bf Periodic problems vs interval problems.} There is a clear similarity between the periodic problems and the interval problems based on the double-row monodromy matrix. Conceptually,  the monodromy matrices in both cases are obtained by integration along a loop, which naturally leads to Lax formulations. For the interval problems, the extra set of constraints \eqref{eq:rmpm} are the key ingredients to determine  admissible classes of boundary conditions and to ensure integrability of the model. Technically, the interval problems can be extended as periodic problems with the length of periodicity being twice of that of the interval as illustrated in Fig.~\ref{fig:monomatrixsddd}. This is based on the following equality (the last equality serves as a definition for $\widetilde{T}_{-L}(\lambda)$)  
  \begin{equation}\label{eq:TTT}
T_L(\lambda) = \overset{\curvearrowleft} {\exp}  \int^{L}_{0} U(\xi;\lambda)d\xi = \overset{\curvearrowleft} {\exp}\int_0^{-L}\widetilde{U}(\xi;\lambda)d\xi =\widetilde{T}_{-L}(\lambda)\,,
\end{equation}
where
\begin{equation}
  \label{eq:10}
\widetilde{U}(x;\lambda): = i\lambda\sigma_3 - Q(-x)\,. 
  \end{equation}
  Therefore,  by preparing an extended Lax matrix $U_{\text{ext}}$ in the form
\begin{equation}\label{eq:qext}
  U_{\text{ext}}(x;\lambda) =
  \left\{   \begin{aligned}  \delta_LK_+(\lambda) \quad & \text{for~} x=L \\U(x;\lambda) \quad & \text{for~} 0<x<L  \\ 
\delta_0K_-(\lambda) \quad & \text{for~} x=0 \\    \widetilde{U}(x;\lambda) \quad  & \text{for~} -L<x<0 \\
     \text{extended } & \text{to be of period}~2L
  \end{aligned}\right.\,,
\end{equation}
with $\delta_L=\delta(x-L) $ and $\delta_0=\delta(x) $ being the Dirac delta functions, one could establish the equivalence between the periodic monodromy matrix for the extended Lax matrix $U_{\text{ext}}$  \eqref{eq:qext} and the double-row monodromy matrix \eqref{eq:drmm1}. Note that this periodic extension is  reminiscent of the method of reflection for solving interval problems for linear PDEs subject to Robin boundary conditions (see, for instance, \cite[Section $3.2$]{SW} as an example). The two particular points $x=L$ and $x=0$ (and their periodic extensions to the whole axis) can be interpreted as ``defect points'' associated with defect conditions determined respectively by the constraints \eqref{eq:k2} and \eqref{eq:k1}. In Section~\ref{sec:44},  explicit examples of this extension for NLS on the half-line  are provided. 

  \item {\bf Involution relation.} The double-row monodromy matrix admits an extra involution relation 
   \begin{equation}\label{eq:gg1}
\Gamma_L^{-1}(\lambda) = K_+(-\lambda)\Gamma_L(-\lambda)K_+(\lambda)\,. 
\end{equation}
This is established by taking account of the involution relation
\begin{equation}
  K_+(-\lambda)K_+(\lambda) = I\,. 
\end{equation}
\item {\bf Interval problems vs half-line problems.} A na\"ive approach to the half-line problems on the positive semi-axis is to consider a single-row transition matrix (here we assume the NLS field belongs to the functional space of Schwartz-type, hence decays rapidly as $x\to \infty$) as 
  \begin{equation}\label{eq:Tinfyt}
    T_{\infty}(\lambda) =   \lim_{L\to \infty}\left(e^{i\lambda L \sigma_3}\, T_L(\lambda)\right) 
    = \bma a(\lambda) & b(\lambda) \\ -b^*(\lambda)& a^*(\lambda) \ema\,, 
    \quad \lambda \in \RR\,,\end{equation}
where $T_L(\lambda) = T(L,0;\lambda)$ is understood.  Then, the two fundamental solutions
  \begin{equation}
    G(x;\lambda) = T(x,0;\lambda)\,,\quad F(x;\lambda) = \lim_{y\to \infty}T(x,y;\lambda)e^{-i\lambda y \sigma_3}\,,
  \end{equation}
  are connected by $T_{\infty}(\lambda)$ as
  \begin{equation}
    G(x;\lambda) =  T_\infty(\lambda)F(x;\lambda)\,.
  \end{equation}
The entries $a(\lambda)$ and $b(\lambda)$ (resp.~$a^*(\lambda^*)$ and $b^*(\lambda^*)$)  are analytic and bounded in the upper (resp.~lower) half complex plane, and  $b(\lambda)$ encodes the initial data on the half-line, {\it cf}.~\cite{fokas2002integrable, fokas2008}. 
However, this transition matrix does not capture the boundary behaviors; no Lax formulation exists  in this case. Based on Sklyanin's formalism by extending one endpoint of the interval to infinity, one can provide an overall satisfactory description of the half-line problems in the presence of a boundary with boundary conditions characterized by the constraint \eqref{eq:kvk}. In particular, the existence of infinity as a reference point in the scattering system enables us to identify Jost solutions as in the case of full line problems.  This also makes the spectral properties of the scattering system related to somehow ``simpler''  mathematics than the case of interval problems, since  the spectral parameter in latter case may live in some compact hyper-elliptic Riemann surface\footnote{This is due to the periodic extension of the interval problems presented here (see \cite{AM1, bel1} for periodic problems for NLS). Similar conclusions were obtained in \cite{BiIts}, where the interval problems for a particular model, that is the defocusing NLS equation subject to Robin boundary conditions at the two endpoints,  was considered.}. 

\item {\bf Integrable boundary conditions.}  Although not explicitly proved, the boundary conditions derived from the constraint \eqref{eq:kvk}  are indeed {\it integrable boundary conditions}, which also makes the associated  interval problems (and half-line problems) integrable.  This aspect is partially justified by the existence of the Lax formulation \eqref{eq:gammat}, and is supported by the present work where an inverse scattering transform on the half-line is developed leading to multi-soliton solutions. A systematic treatment of integrability based on the  Hamiltonian formalism can be found in \cite{ACC}.       
\item {\bf Comparison with Fokas' method.}  The unified transform method, developed by Fokas, can be considered as a generalization of the inverse scattering transform, and   provides a powerful analytic framework to deal with generic initial-boundary value problems for a wide class of integrable PDEs \cite{fokas1997unified, fokas2002integrable, fokas2008}. It relies on simultaneous treatments of the space and time parts of the Lax pair. The initial and boundary data are transformed into scattering systems, which are formulated as certain Riemann-Hilbert problems.  However, in contrast to Sklyanin’s approach,  there is no clear definition of integrable boundary conditions in the unified transform; although asymptotic analysis for large time can be implemented using the nonlinear steepest descent method \cite{DZhou}, it is a difficult task to solve explicitly the Riemann-Hilbert problems. 

\end{itemize}
\section{Inverse scattering transform for half-line problems}\label{sec:4}
In this section, we develop the inverse scattering transform for the NLS equation \eqref{eq:fnls} on the half-line. This relies on extension of the integrable interval problems presented in the previous section by sending one endpoint of the interval to infinity. A fair proportion of the present content is in line with the inverse scattering transform for NLS on the whole line. We refer readers to the monographs \cite{faddeev, NMSP} for details. Due to the presence of a boundary and the associated boundary conditions, extra properties of the scattering systems are needed. The inverse part is formulated by virtue of a Riemann-Hilbert problem. Equivalence between our approach and the nonlinear method of reflection \cite{fokas1989initial, biondini2009solitons} is also provided.

\subsection{Scattering system for half-line problems}
\label{sec:41}We perform what is conventionally called {\em direct scattering transform} for the half-line problems. Without loss of generality, we consider the NLS equation \eqref{eq:fnls} being defined on the positive semi-axis, {\it ie}.~$x> 0$. By assuming that a smooth solution $q(x,t)$ on the half-line exists, the associated scattering system  for the half-line problem is inherited from  that of the interval problems \eqref{eq:drmm11} by extending the endpoint $L$ to infinity. We work with  $K_\pm$ belonging to the hierarchy of reflection matrices as explained in Section~\ref{sec.32}.
We also assume the NLS field belonging to the functional space of  Schwartz-type
\begin{equation}\label{eq:schl}
q(x,t_0) \in \cS(\RR^+)\,, 
\end{equation}at a fixed time $t=t_0$.  This imposes that   $q(x,t_0)$ and its derivatives decay rapidly as $x\to \infty$. In other words,  the vanishing boundary conditions are imposed at infinity.  
\begin{rmk}
This particular choice of the initial data in the sense of \eqref{eq:schl} will make the analysis more straightforward than the case that the $q(x,t_0)$ tends to a constant as $x \to \infty$, which corresponds the half-line NLS model on a constant background. In the later case, the spectral parameter $\lambda$ is living on a two-sheet Riemann surface (see, for instance, \cite{AK1} for the focusing case,  and~\cite{ZS3,  faddeev} for the defocusing case), and the whole analysis could be modified accordingly. Note that there is a lack of investigation for the half-line NLS model on a constant background. This problem may itself  be of interest to many physics applications, and will be reserved for further work.    
\end{rmk}

The boundary conditions at $x=0$ is determined by the reflection matrix $K_-$. Let (we drop the subscript of $\kappa$ for convenience)
\begin{equation}
  \label{eq:K15}
  \det K_-(t;\lambda) = \kappa(\lambda)\,. 
\end{equation}
 For $q(x,t_0)$ in the sense of \eqref{eq:schl},  the time-part Lax matrix of NLS behaves as $\lim_{x\to\infty}V = -2i\lambda^2\sigma_3$.  This allows us to choose the reflection matrix  $K_+$  as a constant diagonal matrix. We fix $K_+(\lambda)$ as
\begin{equation}
  \label{eq:Kpi}
  K_+(\lambda) =  \bma 1 & 0 \\ 0 & \kappa(\lambda) \ema 
\end{equation}
with $\kappa(\lambda)$ being defined in \eqref{eq:K15}. This form of $K_+$ is compatible with the vanishing boundary conditions at infinity. 
It also satisfies the normalization property (R), and admits a semi-discrete Lax  interpretation in the sense of \eqref{eq:KETA}. The latter point is justified in Appendix \ref{ap:0}. It is crucial to keep in mind that, despite being a constant matrix,  the reflection matrix $K_+(\lambda)$ is explicitly present at infinity, and delivers a discrete action on the auxiliary field by transforming $\lambda$ to $-\lambda$. 

To establish the scattering system for the half-line problem, we follow that of the interval problem \eqref{eq:drmm11} by sending $L \to \infty$. In a way similar to  \eqref{eq:1fgfl} where the extended fundamental solutions are defined for the full line problem, we define  two fundamental solutions  $F$ and $G$  in the forms
\begin{subequations}
  \begin{align}
  F(x;\lambda)& = \lim_{L\to\infty} T(x,L;\lambda)\,e^{-i\lambda L \sigma_3}\,,\\   G(x;\lambda)& = \lim_{L\to\infty} T(x,0;\lambda)\,K_-(-\lambda)\,T^{-1}_L(-\lambda)\,e^{i\lambda L \sigma_3}\,K_+(\lambda) \,,
\end{align}
\end{subequations} for $\lambda\in \RR$, where  $T_L(\lambda)= T(L,0;\lambda)$ is understood. 
They play the roles of Jost solutions as in the case of full line problems. Recall the definition of the half-line transition matrix $T_\infty(\lambda)$ in \eqref{eq:Tinfyt}, then  $T^{-1}_\infty(\lambda) = \lim_{L\to \infty} T^{-1}_L(\lambda)\,e^{-i\lambda L \sigma_3}$, and one has $F, G$ satisfying the following boundary/asymptotic behaviors
\begin{subequations}
\begin{align}
F(0;\lambda) = T^{-1}_\infty(\lambda)\,, \quad &  \lim_{x\to \infty}F(x;\lambda) = e^{ -i \lambda x\sigma_3}\,,  \\
G(0;\lambda) =  H(\lambda)\,, \quad &   \lim_{x\to \infty}G(x;\lambda) = e^{ -i \lambda x\sigma_3}T_\infty(\lambda)\,H(\lambda)\,,
\end{align}     
\end{subequations}
where the quantity $H(\lambda)$ is defined as
\begin{equation}
H(\lambda):=  K_-(-\lambda)\, T^{-1}_\infty(-\lambda)\,K_+(\lambda) \,. 
\end{equation}
Clearly, $F$ and $G$ are connected by an extended (half-line) monodromy matrx $\Gamma(\lambda)$ as 
\begin{equation}
  G(x;\lambda)= F(x;\lambda)\,\Gamma(\lambda)\,,
\end{equation}
where
\begin{equation}\label{GGGG}
  \Gamma(\lambda) = T_\infty(\lambda)\,K_-(-\lambda)\, T^{-1}_\infty(-\lambda)\,K_+(\lambda) \,. 
\end{equation}

The particular form of $K_+$  \eqref{eq:Kpi} imposed at infinity ensures $
   \det \Gamma(\lambda)    = 1$. 
 Following the transformations
\begin{equation}
X(x;\lambda) =  F(x;\lambda)  e^{i \lambda x\sigma_3}  \,,\quad Y(x;\lambda) =  G(x;\lambda)  e^{ i \lambda x\sigma_3}\,,
\end{equation}
the matrix-valued functions $X$ and $Y$  admit the  integral representations
\begin{subequations}
\begin{align}
  X(x;\lambda) =&I+\int_{\infty}^xe^{-i\lambda(x-\xi)\sigma_3}Q(\xi)X(\xi;\lambda)e^{i\lambda(x-\xi)\sigma_3}d\xi\,,\\
  Y(x;\lambda) =&e^{-i\lambda x\sigma_3}H(\lambda) e^{i\lambda x\sigma_3}+\int_{0}^xe^{-i\lambda (x-\xi)\sigma_3}Q(\xi)Y(\xi;\lambda)e^{i\lambda(x-\xi)\sigma_3}d\xi\,. 
\end{align}
\end{subequations}  
They are connected by the double-row monodromy matrix $\Gamma(\lambda)$ as \begin{equation}
  \label{eq:YXG}
  Y (x;\lambda) =  X(x;\lambda) e^{-i\lambda x \sigma_3}\Gamma(\lambda) e^{i\lambda x \sigma_3}\,,\quad \lambda \in \RR\,. 
\end{equation}

It follows from  standard techniques in integral functions that $X$ and $Y$, as functions of the spectral parameter $\lambda$,  can be split into column vectors with different domains of analyticity in the complex plane.
Some extra care is needed for the function $Y$. Due to the presence of the reflection matrices $K_\pm$, one can write down $Y(0;\lambda)$ as
\begin{equation}
  \label{eq:20YY}
Y(0;\lambda) = H(\lambda) =\bma K_{11} a^*(-\lambda)  + K_{12} b^*(-\lambda) & \kappa(\lambda)(K_{12} a(-\lambda) -K_{11} b(-\lambda)  ) \\ K_{21}  a^*(-\lambda) + K_{22} b^*(-\lambda) & \kappa(\lambda) (K_{22}a(-\lambda) -K_{21} b(-\lambda) )\ema  , 
\end{equation}
for $\lambda\in \RR$,  where $K_{ij}$, $1\leq i,j\leq 2$, denote the $(ij)$ entries of $K_-(-\lambda)$, and $a(\lambda)$, $b(\lambda)$ are the entries of the half-line transition matrix $T_\infty(\lambda) $ defined in \eqref{eq:Tinfyt}.

Recall the definition of $\kappa(\lambda)$ \eqref{eq:normk1}, \ie $\kappa(\lambda) = \epsilon f(\lambda)/g(\lambda)$,  and the form of reflection matrix \eqref{eq:LKa}, \ie $K(\lambda)= \cL(\lambda)/g(\lambda)$. The entries of the reflection matrix $K_-(-\lambda)$ have simple poles for $\lambda$ belonging to $ \fD$ that is the set of zeros of $f(\lambda)$, since $g(-\lambda) =  (-1)^\cM f(\lambda)$.  Also recall that  the functions $a(\lambda), b(\lambda)$ are analytic and bounded in the upper half complex plane $\CC^{(+)}$, {\it cf}.~\cite{fokas2002integrable, fokas2008}. Here, we make  further requirements for $\kappa(\lambda)$ that all points in $\fD$ are strictly located in the lower half complex plane $\CC^{(-)}$.  By doing so,   the first column vector of $Y(0;\lambda)$, as a function of $\lambda$,  is analytic and bounded in $\CC^{(+)}$.  Similarly, the second column vector of $Y(0;\lambda)$ is analytic and bounded in  $\CC^{(-)}$. This leads to the analytic properties of $X$ and $Y$.

\begin{lemma}\label{lem:xy}
Let all points of $\fD$, that are zeros of $\kappa(\lambda)$,  be strictly located in the lower half complex plane $\CC^{(-)}$. Then, the functions $X$ and $Y$,  as functions of the spectral parameter $\lambda$,   can be written as
  \begin{equation}
X = (X^{(-)}, X^{(+)})\,,\quad Y = (Y^{(+)}, Y^{(-)})\,. 
\end{equation}
The column vectors $X^{(+)}, Y^{(+)}$ (resp. $X^{(-)}, Y^{(-)}$) are analytic and bounded in $\CC^{(+)}$ (resp.  $\CC^{(-)}$). 
\end{lemma}

  In contract to the assumption on $\fD$ we made in Lemma~\ref{lem:xy},  the general definition  of $\kappa(\lambda)$ in Section \ref{sec.32} does not require the points of  $\fD$ to be located in $\CC^{(-)}$. However, for the half-line problem we consider here, this assumption allows us to avoid {\em unnecessary} singularities of $Y^{(+)}$ (resp.  $Y^{(-)}$) in $\CC^{(+)}$ (resp.  $\CC^{(-)}$). In fact, following the definition  \eqref{eq:normk1},   we can always rearrange the zeros and poles of $\kappa(\lambda)$ so that all zeros of  $\kappa(\lambda)$  are located in $\CC^{(-)}$. In the other extreme case that all points of $\fD$ are located in $\CC^{(+)}$,  $Y^{(+)}$ (resp. $Y^{(-)}$) has simple poles when $\lambda$ takes value in $\fD$ (resp. in $\fD^*$) due to the presence of the reflection matrices $K_\pm$. Then,  Lemma~\ref{lem:xy} could  be  accordingly modified: let all points of $\fD$ belong to $\CC^{(+)}$, then  $Y^{(+)}$ (resp. $ Y^{(-)}$) is analytic and bounded in $\CC^{(+)}$ (resp.  $\CC^{(-)}$) except for finite number of points when $\lambda$ takes value in $ \fD$ (resp. in $\fD^*$).

\subsection{Analytic properties of the scattering functions}\label{sec:42}
Having the scattering system \eqref{eq:YXG} and the associated analytic properties stated in Lemma~\ref{lem:xy}, we can conclude a series of analytic properties of the scattering functions that are the entries of the double-row monodromy matrix $\Gamma(\lambda)$.  We assume all points of $\fD$ are located in $\CC^{(-)}$ for simplicity of the presentation. More generic situation is discussed in Remark~\ref{rmk:gc}.

\begin{itemize}
\item For $\lambda\in \RR$, the double-row monodromy matrix  $\Gamma$ can be expressed as  \begin{equation}\label{eq:ABCD}
    \Gamma(\lambda) = 
    \bma A(\lambda) & -{B}^*({\lambda^*}) \\ B(\lambda) & {A}^*({\lambda^*})\ema
  \,, 
\end{equation}
for $\lambda\in \RR$, which is a direct consequence of the the involution relations  \eqref{eq:inv1} and the normalization property  (R). The scattering system \eqref{eq:YXG} yields
\begin{subequations}\begin{align}
\label{eq:AA12}  A(\lambda) =\det (Y^{(+)}, X^{(+)})\,,\quad   &  A^*(\lambda^*) =\det (X^{(-)}, Y^{(-)})\,, \\
\label{eq:BB1}  B(\lambda) =\det (X^{(-)}, Y^{(+)})\,,\quad   & B^*(\lambda^*) =\det (X^{(+)}, Y^{(-)})\,,  \end{align}
\end{subequations}  
which means $A(\lambda)$ (resp. $A^*(\lambda^*)$) is analytic and bounded in $\CC^{(+)}$ (resp. $\CC^{(-)}$). \item The scattering function $B(\lambda)$ plays the role of {\em reflection coefficient} for the  scattering system \eqref{eq:YXG}, and, in general, can only be well-defined for $\lambda\in\RR$.  For smooth initial data $q(x,t_0)$ in the sense of \eqref{eq:schl}, $B(\lambda)$ is also a smooth function encoding both the initial data and the boundary conditions\footnote{This can be seen by explicitly expanding $B(\lambda)  $ as (using \eqref{eq:BB1} evaluated at $x = 0$) \[B(\lambda) =    a^*(\lambda)\left[K_{21}  a^*(-\lambda) + K_{22} b^*(-\lambda)\right] -b^*(\lambda) \left[K_{11} a^*(-\lambda)  + K_{12} b^*(-\lambda)   \right]\,,  \]where $a(\lambda), b(\lambda)$ are entries of $T_\infty(\lambda)$ defined in \eqref{eq:Tinfyt}, and $K_{ij}$, $1\leq i,j\leq 2$, denote the $(ij)$ entries of $K_-(-\lambda)$. Then,  $B(\lambda)$ involves both $b(\lambda)$, which encodes the initial data on the half-line, and the entries of $K_-(-\lambda)$, which contain information of the boundary conditions. }. 
 Due to $\det \Gamma(\lambda) =1$, one has
\begin{equation}\label{eq:AA1}
  A(\lambda)A^*(\lambda) = 1- B(\lambda)B^*(\lambda)\,,\quad \lambda\in \RR\,.
\end{equation}
This is a multiplicative Riemann-Hilbert problem with the asymptotic behaviors
\begin{equation}
  \label{eq:21AA}
\lim_{\lambda\to \infty}A(\lambda) = 1 + O(1/\lambda)\,,\quad \lim_{\lambda\to \infty}B(\lambda)  = O(1/\lambda)\,, 
\end{equation}
which can be derived  by taking account of the asymptotic behaviors of $a(\lambda)$ and $ b(\lambda)$, {\it cf}.~\cite{fokas2002integrable, fokas2008} as well as the requirement  \eqref{eq:asymK}. 
\item The involution property \eqref{eq:gg1} implies that
\begin{equation}\label{eq:aafbb}
  A(\lambda) = {A}^*(-{\lambda})\,,\quad B(\lambda) = -\kappa(-\lambda)B(-\lambda)\,, 
\end{equation}
for $\lambda\in \RR$. This relation is crucial to the characterization of the half-line problem.  \item It follows from Lemma \ref{lem:kkpm} and  \eqref{GGGG}  that 
  \begin{equation}
    \label{eq:19}
\frac{d}{dt}    \Gamma(t;\lambda)  = \lim_{L\to\infty} [V(L,t;\lambda), \Gamma(t;\lambda)] = -2i\lambda^2[ \sigma_3, \Gamma(t;\lambda)]\,.
\end{equation}
This corresponds to the {\em time evolution} of the scattering functions.  In components,  $A$ is preserved by
the time evolution, and  $B(t;\lambda)= B(0;\lambda)e^{4i \lambda^2t}$. 
\end{itemize}

\begin{figure}[ht]
  \centering
  \begin{tikzpicture}[scale=.9]
    \def\l{3}%
    \tikzstyle{nod1}= [circle, inner sep=0pt, fill=white, minimum size=4pt, draw]

    \coordinate (p1) at (-\l,0);
    \coordinate (p2) at (\l,0);
    \coordinate (q1) at (0,-\l);
    \coordinate (q2) at (0,\l);
    \draw[->] (p1)  -- (p2);
    \draw[->] (q1)  -- (q2) ;
    \node at (\l*0.85, \l*0.75) {$\CC^{(+)}$};
    \node at (\l*0.85, -\l*0.75) {$\CC^{(-)}$};
    
    \node[nod1] (r1) at (\l/2,\l/2) [label= above right:$r_j$ ] {};
    \node[nod1] (tr1) at (-\l/2,\l/2) [label= above left:$-r_j^*$ ] {};
    \draw[ dotted] (r1) -- (tr1);
    \node[nod1] (b1) at (\l/2,-\l/2) [label= below right:$r^*_j$ ] {};
    \node[nod1] (tb1) at (-\l/2,-\l/2) [label= below left:$-r_j$ ] {};
    \draw[ dotted] (b1) -- (tb1);
    
    \node[nod1] (ap) at (0,\l/3) [label= left:$is_\ell$ ] {};
    \node[nod1] (ap) at (0,-\l/3) [label= left:$-is_\ell$ ] {};
  \end{tikzpicture}
  \caption{Zeros of  $A(\lambda)$ in $\CC^{(+)}$ and $A^*(\lambda)$ in $\CC^{(-)}$   in the presence of integrable boundary conditions. } \label{fig:51011}
\end{figure}
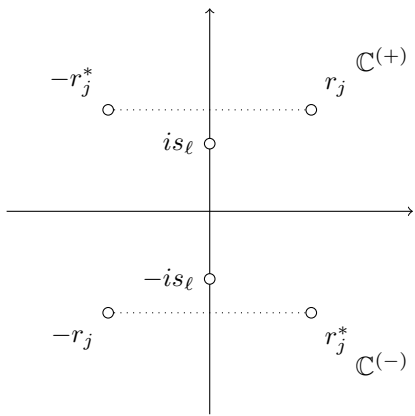

The zeros of the scattering function $A(\lambda)$ in $\CC^{(+)}$  are associated to ``bound states'' of the space-part of the Lax equations. To simplify our analysis, we assume that  $A(\lambda)$ has only finite number of simple zeros  in $ \CC^{(+)}$.  Moreover, the zeros of $A(\lambda)$  do not coincide with any point of $\fD^*$ (note that the points of $\fD$ are required to be located in $\CC^{(-)}$, so that the points of $\fD^*$ are located in  $\CC^{(+)}$). 
Let $r_j \in \CC^{(+)}$ be a complex number with non zero real part, \ie $\text{Re}\, r_j \neq 0$ and $\text{Im}\, r_j > 0$.  If $A(r_j) = 0$, then  $A(-r_j^*) = 0$.  This is because  $A(r_j) = 0$ implies $A^*(r^*_j) = 0$; using the first involution property in \eqref{eq:aafbb} yields $A(-r_j^*) = 0$ and $A^*(-r_j) = 0$. Therefore, $r_j$ is simultaneously paired with $-r_j^*$ as zeros of $A(\lambda)$. This pairing mechanism  of zeros is illustrated in Fig.~\ref{fig:51011}. In the special case that a zero of $A(\lambda)$ is a pure imaginary number $is_\ell$, for $s_\ell >0$, the  paired zero coincides with itself. The pure imaginary zero of $A(\lambda)$ is set to be simple according to our assumption.

By taking account of the above consideration, one introduces a presumed rational function $\zeta(\lambda)$  describing zeros of $A(\lambda)$
\begin{equation}
  \label{eq:zeta12}
  \zeta(\lambda) =\prod_{j=1}^N\left[\frac{\lambda-r_j}{\lambda-r_j^*}\frac{\lambda+r^*_j}{\lambda+r_j}\right]\prod_{\ell=1}^M  \left[\frac{\lambda-i s_\ell}{\lambda+is_\ell}\right]\,, 
\end{equation}
where $r_j$, $j =1,\dots,N$, are distinct points with non zero real parts strictly located in $\CC^{(+)}$, and $s_\ell$,  $\ell =1,\dots,M$, are distinct real positive numbers. These points do not coincide with any point of $\fD^*$.  Under this assumption, one could express $A(\lambda)$ in terms of the reflection coefficient $B(\lambda)$.
 \begin{lemma}\label{lemma:1}
Consider the presumed factor \eqref{eq:zeta12}.  Then,  $ A(\lambda)$ can be expressed as 
  \begin{equation}
A(\lambda) = \zeta(\lambda)
\exp\left[\frac{1}{i\pi}\int_{ 0  }^{\infty}\frac{\lambda \log[1 -|B(\mu)|^2]}{\mu^2-\lambda^2}d\mu\right] \,,  \end{equation}
for $\lambda \in \CC^{(+)}$. The quantity $\log A(\lambda)$ admits the following expansion
\begin{equation}
  \label{eq:12la}
  \log A(\lambda) = \sum_{j=0}^\infty \frac{I_{2j}}{(2i \lambda)^{2j+1}}\,,
\end{equation}
where 
  \begin{equation}\label{eq:tr1}
        I_{2j}=-\frac{2}{\pi}\int_0^\infty \log \left[1-|B(\mu)|^2\right](2i \mu)^{2j}\,d\mu+2 \sum_{\ell=1}^N \frac{(2ir^*_\ell)^{2j+1}-(2ir_{\ell})^{2j+1}}{2j+1}+2\sum_{\ell=1}^{M}  \frac{ (2s_\ell)^{2j+1}}{2j+1}\,.
  \end{equation}

\end{lemma}
The above formulae are known as {\em trace identities}. The proof is provided in Appendix \ref{ap:1}. The quantities $I_{2j}$, $j = 0,1,2,\dots$, are related to the conserved quantities of the model, and  the index  increments by $2$  due to the presence of a boundary. This is in contrast to the full line problem for NLS\footnote{For instance, in the case of a full line problem, the quantity $I_1$, which is associated with the translation symmetry, is broken for the half-line problem. }. Similar results were obtained, for instance in \cite{BT1}, for the case of NLS on the half-line subject to Robin boundary conditions. 

In the case  $A(\lambda)\vert_{\lambda =r_j} =0 $,  the column vector $  X^{(+)}(r_j)$ is proportional to $  Y^{(+)}(r_j)$. This means the first column vector of $G$ and second column vector of $F$ are bound solutions of the space-part of the Lax equations \eqref{eq:UVee}. Let $\gamma_j$ be the proportionality coefficient
\begin{equation}
  \label{eq:21XY}
  X^{(+)}(r_j) = \gamma_j\, Y^{(+)}(r_j)\,, \quad 0<|\gamma_j|<\infty\,,  
\end{equation}
evaluated at $x = 0$. Such  $\gamma_j$ is known as {\em  norming constant} associated with the discrete spectrum $r_j$. When working in the class of exponentially fast decaying initial data, there is a nice characterization of  $\gamma_j$ in terms of $B(\lambda)$,  which is assumed to be able to be analytically extended off the real axis so that $B(r_j)$ is finite. Then, it follows from \eqref{eq:BB1} that $\gamma_j$ can be expressed as
\begin{equation}
  \label{eq:26}
\gamma_j =B^*(r_j^*) =\frac{1}{ B(r_j)} \,. 
\end{equation}
For a paired zero $-r_j^*$ of $A(\lambda)$, the associated norming constant $\widetilde{\gamma}_j$ is paired with $\gamma_j$ as
\begin{equation}\label{eq:ggr}
  \widetilde{\gamma}^*_j\gamma_j = -\kappa(r_j)\,,
  \end{equation}due to the involution relation \eqref{eq:aafbb}. 
The paired norming constants $\{\gamma_j, \widetilde{\gamma}_j\}\vert_{j=1,\dots,N}$ evolve with the space and time variables as \begin{equation}
\gamma_j (x,t) = \gamma_j e^{-2i\theta(x,t;r_j)}
\,,\quad \widetilde{\gamma}_j (x,t) = \widetilde{\gamma}_j e^{-2i\theta(x,t;\widetilde{r}_j)}\,, 
\end{equation}
where $\theta(x,t;\lambda) = \lambda(x+ 2 \lambda\, t)$ and $\widetilde{r}_j = -r_j^*$.

In the case that $A(\lambda)$ has a pure imaginary zero, \ie $A(is_\ell) = 0$, let $\eta_\ell$ be the associated norming constant. Again due to the involution relation \eqref{eq:aafbb}, one has
\begin{equation}
  \label{eq:12eta}
  {\eta}^*_\ell\eta_\ell = |\eta_\ell|^2 = -\kappa(is_\ell)\,.
\end{equation}
This means that the quantity $-\kappa(is_\ell)$ must be a positive real number, which makes further restriction on the choice of $s_\ell$. The above relations conclude the characterization of the so-called discrete scattering data
\begin{equation}
  \label{eq:28dsd}
\{r_j,\widetilde{r}_j; \gamma_j, \widetilde{\gamma}_j\}_{j=1,\dots,N} \cup \{is_\ell; \eta_\ell\}_{j=1,\dots,M}\,, 
\end{equation}
where $\widetilde{r}_j = -r_j^*$,  $\gamma_j$ and $\widetilde{\gamma}_j$ are related by the constraint \eqref{eq:ggr}, and $\eta_\ell$ is characterized by \eqref{eq:12eta} with $\kappa(is_\ell)< 0$.  In the reflectionless case, \ie $B(\lambda) = 0$ for $\lambda \in\RR$,  $A(\lambda)  = \zeta(\lambda)$, the set of scattering data $ \{r_j,-r_j^*; \gamma_j, \widetilde{\gamma}_j\}_{j=1,\dots,N} $  will generate $N$  ``moving'' solitons on the half-line, and $ \{is_\ell; \eta_\ell\}_{j=1,\dots,M}$  will generate $M$ ``static'' solitons bounded to the boundary. Overall, they form  $(N+M)$-soliton solutions on the half-line.

Here, we discuss possible behaviors of $A(\lambda)$ when $\lambda$ takes value in   $\fD^*$. In general, we assume   $A(\lambda) \neq 0$ for $\lambda\in \fD^*$.  In the special case that  $A(\lambda) $ has a simple zero for $\lambda\in \fD^*$, assuming there exists a norming constant $\gamma_j$ in the sense of \eqref{eq:21XY}, then the paired norming constant $\widetilde{\gamma}_j$ according to \eqref{eq:ggr} tends to   $ \infty $ which contradicts the assumption on the norming constants. This case is excluded.  If $X^{(+)} = 0$ and/or $Y^{(+)} = 0$ for $\lambda \in \fD^*$, then they are associated to degenerate states. 

\begin{rmk}
  \label{rmk:gc}
  Instead of assuming all points of $\fD$ are located in $\CC^{(-)}$, consider the other extreme case that all points of $\fD$ are located in $\CC^{(+)}$. Then, $Y^{(+)}$ as well as $A(\lambda)$ has  finite number of simple poles when $\lambda$ takes value in $\fD$, and one can show that  $A(\lambda)$ is expressed as
  \begin{equation}
    \label{eq:6}
    A(\lambda) = \epsilon \kappa(-\lambda)\zeta(\lambda)\widetilde{A}(\lambda)\,,  \end{equation}
for $\lambda\in\CC^{(+)}$, with   $\widetilde{A}(\lambda)$ being defined in \eqref{eq:tAB}. One adds  the parity parameter $\epsilon$ in the above expression to make the normalization $\lim_{\lambda\to\infty}A(\lambda) =1$ valid.    In this case, the scattering system \eqref{eq:YXG} can be defined on a modified complex plane by excluding all points belonging to $\fD\cup \fD^*$, and  analytic properties of the scattering functions follow accordingly.  Note that this case is in line with the non-degeneracy assumption of the reflection matrices made in Lemma~\ref{lem:kkpm}.
\end{rmk}

\subsection{Riemann-Hilbert formulation of the NLS equation on the half-line }\label{sec:43}
The above subsection completes the direct scattering transform for NLS equation on the  half-line subject to integrable boundary conditions determined by a reflection matrix $K_-(t; \lambda)$ associated with a given $\kappa(\lambda)$ in the sense of \eqref{eq:normk1}. The inverse part is formulated in this section as a matrix Riemann-Hilbert problem. This sets up a generic analytic framework for the half-line problem under consideration. Possible solutions of the matrix Riemann-Hilbert problem inversely solve the NLS equation on the half-line. 

Let us first state the Riemann-Hilbert problem inherited from the scattering systems  \eqref{eq:YXG}. We refer readers to \cite{faddeev} for the construction of the Riemann-Hilbert problem. Note that the space variable $x$, as a parameter in the Riemann-Hilbert problem, is restricted to the positive semi-axis, \ie $x>0$.
\begin{enumerate}
\item 
 Let two $2\times 2$ matrix-valued function $J^{(+)}(x,t;\lambda)$ and $J^{(-)}(x,t;\lambda)$ in the forms
 \begin{equation}\label{JJJ}
   J^{(+)} = (Y^{(+)},X^{(+)})\,,\quad    J^{(-)}= \det(X^{(-)},Y^{(-)})(X^{(-)},Y^{(-)})^{-1}\,. 
 \end{equation}
 It follows from Lemma~\ref{lem:xy} that $J^{(+)}$ (resp.  $J^{(-)}$), as a function of $\lambda$, is analytic and bounded in $\CC^{(+)}$ (resp. $\CC^{(-)}$). By definition,
 \begin{equation}
   \label{eq:20AA}
\det   J^{(+)} = A(\lambda)\,,\quad \det   J^{(-)} = A^*(\lambda)\,.
 \end{equation}

 \item $J^{(+)}$ and  $J^{(-)}$ are connected by  the  following jump condition  \begin{equation}
\label{JJCDN}J^{(-)}J^{(+)} = e^{-i \theta(x,t;\lambda)\sigma_3}\,J(\lambda)\,e^{ i\theta(x,t;\lambda)\sigma_3 }\,,\quad \theta(x,t;\lambda) :=  \lambda(x+ 2 \lambda\, t)\,,
\end{equation}
for  $\lambda\in \RR$. The jump matrix $J(\lambda)$ is in the form
\begin{equation}
\label{jj}
J(\lambda) =  I + \widetilde{J}(\lambda) 
\,,\quad \widetilde{J}(\lambda) = \bma 0 & {B}^*({\lambda}) \\ B(\lambda)&0\ema \,. 
\end{equation}
\item By assuming the initial data $q(x,t_0)$ to be  a smooth function in the sense of \eqref{eq:schl},  the reflection coefficient $B(\lambda)$ is smooth for $\lambda \in \RR$ obeying   \begin{equation}
\label{sym}    \kappa(\lambda)B(\lambda) = -B(-\lambda) \,, 
\end{equation}
for $\kappa(\lambda)$ being in the form of \eqref{eq:normk1} with its zeros strictly located in $\CC^{(-)}$. 
\item Under the assumption that $A(\lambda)$ has a finite number of simple zeros strictly located in $\CC^{(+)}$, one has
\begin{equation}
\label{sym1}
 |B(\lambda)|<1\,, \quad \lambda\in \RR\,.  
\end{equation}
 
\item $J^{(\pm)}$ obey the  normalization properties
 \begin{equation}\label{eq:qI}
   \lim_{\lambda\to \infty}    J^{(\pm)} =I\,. 
 \end{equation}
\end{enumerate}
Apart from the extra involution condition \eqref{sym} for the reflection coefficient $B(\lambda)$,  the above matrix Riemann-Hilbert problem for NLS on the half-line is in the same form as in the case of NLS on the whole line.  The next step is to establish inversely the equivalence between the Riemann-Hilbert problem and the half-line NLS model under consideration.

\begin{theorem}\label{th:11}Consider the Riemann-Hilbert problem described by the jump condition (\ref{JJCDN}-\ref{jj}). $J^{(\pm)}$, as functions of $\lambda$, are analytic and bounded in $\CC^{(\pm)}$, and obey the  the normalization properties \eqref{eq:qI}. Moreover, we assume  $B(\lambda)$ is a smooth function for $\lambda\in \RR$  obeying \eqref{sym1}  as well as the involution relation \eqref{sym} for a given $\kappa(\lambda)$ in the sense of \eqref{eq:normk1} with its zeros strictly located in $\CC^{(-)}$.  Then, this Riemann-Hilbert problem admits a unique regular solution, \ie $\det J^{(\pm)} \neq 0$, and the term
\begin{equation}\label{qlj}
q(x,t)=\lim_{\lambda\to \infty} 2 i (J^{(+)})_{12}\,, \quad x>0\,,
\end{equation}
with $(J^{(+)})_{12}$ denoting the $(12)$ entry of $J^{(+)}$, solves the NLS equation \eqref{eq:fnls} on the half-line subject to the boundary conditions at $x =0$  determined by
\begin{equation}
    \label{eq:KK12}    \frac{d}{dt}\cK(t;\lambda) = \cV(0,t;-\lambda)\cK(t;\lambda)-\cK(t;\lambda)\cV(0,t;\lambda)\,,
  \end{equation}
  where $\cK$ is a $2\times 2$ matrix obeying the normalization 
  \begin{equation}
    \label{eq:KK13}
\det \cK(t;\lambda) = \kappa(\lambda) \,, \quad    \lim_{\lambda\to \infty} \cK= \bma 1 & 0 \\ 0 & \epsilon \ema\,,  
\end{equation}
and $\cV$ is in the same form as the time-part Lax matrix of NLS with $q$ defined in \eqref{qlj}. 
\end{theorem}
\prf We split the proof into three parts.

Part 1):  existence and uniqueness of regular solutions of the matrix Riemann-Hilbert problem. The particular Riemann-Hilbert problem we consider here actually admits a unique regular solution. We refer readers to \cite{AFF} and reference therein for details.  A unique regular solution $\widehat{J}^{(+)}$, denoting the inverse of ${J}^{(+)}$, can be expressed as  a Fredholm-type integral equation 
\begin{equation}\label{rs}
  \widehat{J}^{(+)}(x,t;\lambda) =  I - \frac{1}{2\pi i}\int_{-\infty}^{\infty}e^{-i\theta(x,t;\mu)\sigma_3}\,\widetilde{J}(\mu)\,e^{i\theta(x,t;\mu)\sigma_3}\,\widehat{J}^{(+)}(x,t;\mu)\frac{d\mu}{\mu-\lambda}\,,
\end{equation}
for $\lambda \in \CC^{(+)}$. This formula can be obtained by rearranging the jump condition \eqref{JJCDN} in the form
\begin{equation}
  \label{eq:20jj}
\left(\widehat{J}^{(+)}-I\right) - \left( J^{(-)}-I\right) = - e^{-i\theta(x,t;\lambda)\sigma_3}\,\widetilde{J}(\lambda)\,e^{i\theta(x,t;\lambda)\sigma_3}\widehat{J}^{(+)}\,.
\end{equation}
By taking account of the normalization properties \eqref{eq:qI},   one recovers \eqref{rs} by applying the Cauchy integral operator.

Part 2): the reconstructed function $q(x,t)$  \eqref{qlj} for $x>0$ solves the NLS equation. Equivalently, one could show that a regular solution $J^{(+)}$ satisfies the linear systems
  \begin{equation}
    \label{eq:4JJJJ}
J^{(+)} _x = \cU J^{(+)}  + i\lambda J^{(+)}\sigma_3\,,\quad J^{(+)} _t = \cV J^{(+)}  + 2i\lambda^2 J^{(+)}\sigma_3\,,\quad 
\end{equation}
where $\cU$ and $\cV$ are in the same forms as the Lax matrices of NLS defined in \eqref{eq:UVee} with $q$ defined in  \eqref{qlj}. This part of the proof relies on the existence and uniqueness of regular solutions of the Riemann-Hilbert problem, and involves standard techniques in integrable systems known as  {\em dressing methods}. We refer readers to \cite{babelon, faddeev} for details.

Part 3): the boundary conditions of the reconstructed function $q(x,t)$ at $x=0$ are determined by (\ref{eq:KK12}-\ref{eq:KK13}). In other words, the matrix $\cK$ belongs to the hierarchy of reflection matrices, so do the associated boundary conditions  (see Section~\ref{sec.32}). 

Consider the jump condition \eqref{JJCDN} as $x\to0$. It follows from the involution property \eqref{sym} of $B(\lambda)$ that the jump matrix $J(\lambda)$ admits an involution relation
  \begin{equation}
e^{ i2\lambda^2 t\sigma_3 }     J^*(-\lambda) e^{- i2\lambda^2 t\sigma_3 }     = \sigma_2 K_+(\lambda)e^{ -i2\lambda^2 t\sigma_3 }J(\lambda)e^{ i2\lambda^2 t\sigma_3 }K_+(-\lambda)\sigma_2\,,
\end{equation}
for $\lambda\in\RR$, where $K_+(\lambda)$ is in the form \eqref{eq:Kpi} for a given $\kappa(\lambda)$. 
This allows us to rearrange the jump condition in the form
  \begin{equation}
    \left[K_+(-\lambda)\sigma_2\left( J^{(-)}(-\lambda)\right)^*\right]\left[    \left( J^{(+)}(-\lambda)\right)^*\sigma_2K_+(\lambda)\right] =e^{ -i2\lambda^2 t\sigma_3 }J(\lambda)e^{ i2\lambda^2 t\sigma_3 }\,.
  \end{equation}
Introduce  two new  function  $S^{(+)}$ and  $S^{(-)}$ as
  \begin{subequations}
  \begin{align}\label{S11}
    S^{(+)} (t;\lambda)&=  \cK(t;-\lambda)\sigma_2\left(J^{(+)}(-\lambda)\right)^*\sigma_2K_+(\lambda)\,,\\
    S^{(-)} (t;\lambda)& =     K_+(-\lambda)\sigma_2\left( J^{(-)}(-\lambda)\right)^*\sigma_2 \cK(t,\lambda)\,,
  \end{align}    
  \end{subequations}
  where the unknown matrix $\cK(t;\lambda)$ is required to obey
  \begin{equation}
\cK(t;\lambda)\cK(t;-\lambda) = I\,,\quad    \lim_{\lambda\to \infty}\cK(t;\pm\lambda) 
=    \bma 1 & 0 \\ 0 & \epsilon \ema \,,  \quad   \det \cK(\lambda) = \kappa(\lambda)\,. 
\end{equation}
Then,  one has 
\begin{equation}
  \label{eq:27}
\lim_{\lambda\to\infty}  S^{(+)} = \lim_{\lambda\to\infty}  J^{(+)} = I, \quad \det S^{(+)} = \det J^{(+)}\,,
\end{equation}
and  $S^{(+)}$ is analytic and bounded in $\CC^{(+)}$. Similarly,   $S^{(-)}$ is analytic and bounded in $\CC^{(-)}$ with the same normalization as $J^{(-)}$.   
By the uniqueness of regular solutions of the Riemann-Hilbert problem, one has
\begin{equation}
  \label{eq:29SJ}
S^{(\pm)} = J^{(\pm)}\vert_{x = 0}\,.
\end{equation}
  Therefore, the quantities $S^{(\pm)} e^{-2i\lambda^2t\sigma_3}$ obey the time-part of the Lax equation for NLS as stated in Part 2), \ie 
  \begin{equation}
    \frac{d }{d t}\left[ S^{(+)}(t;\lambda) e^{-2i\lambda^2t\sigma_3} \right]=  \cV(0,t;\lambda)\left[S^{(+)}(t;\lambda) e^{-2i\lambda^2t\sigma_3}\right].
  \end{equation}
Combining this with \eqref{S11}, and also by taking account of \eqref{eq:4JJJJ},  one can show that  $\cK(t;\lambda)$ and $\cV(0,t;\lambda)$ obey the constraint \eqref{eq:KK12}. 
This completes the proof.
\finprf
\medskip

Once a regular solution is known, one could construct singular solutions of the Riemann-Hilbert problem by adding ``zeros'' to the regular solution. This procedure is known as {\em dressing transformation}, {\it cf}.~\cite{babelon}. In the special case  $B(\lambda) =0$ for $\lambda\in \RR$, the unique regular solution is the identity matrix $I$. One can associated the discrete scattering data \eqref{eq:28dsd} with  the singular solutions. Then, the reconstructed function \eqref{qlj} corresponds to pure soliton solutions of NLS on the half-line. Explicit examples of half-line solitons subject to integrable boundary conditions are  provided in Section \ref{sec:5}. 
Note that given a solution of the Riemann-Hilbert problem, using formula \eqref{eq:29SJ}, one can reconstruct the reflection matrix $\cK(t;\lambda)$ present at $x=0$.   
\begin{corollary}\label{cor:11}
  Let $J^{(+)}$ be a solution of the Riemann-Hilbert problem stated in Theorem~\ref{th:11}. Then, the reflection matrix $\cK$ can be reconstructed as
  \begin{equation}
    \label{eq:12KJK}
    \cK(t;\lambda) =J^{(+)}(0,t;-\lambda)K_+(\lambda)\sigma_2 \left(\widehat{J}^{(+)}(0,t;\lambda)\right)^*\sigma_2 ,
  \end{equation}
  where  $\widehat{J}^{(+)}$ denots the inverse of ${J}^{(+)}$. 
\end{corollary}
This follows directly from  formula \eqref{eq:29SJ} in the proof of Theorem~\ref{th:11}. The reflection matrix $\cK$ present at $x=0$ can be understood as a {\em dressed} form of the constant diagonal reflection matrix   $K_+$  present at infinity. In particular, the reconstructed reflection matrix $\cK(t;\lambda)$  allows us to express explicitly the additional fields associated with the cases of implicit boundary conditions, {\it cf}.~$\cB_2$ in Case $5$ and $\cB_3$ in Case $6$ presented in Section~\ref{sec.32}, by letting $\lambda = 0$. 

\subsection{Connection to the nonlinear method of reflection}\label{sec:44}
The NLS equation on the half-line subject to Robin boundary conditions \eqref{eq:Rb1} has been extensively studied in the literature. In our setting, this model belongs to a particular case of a class of integrable half-line NLS models.  In \cite{fokas1989initial ,  biondini2009solitons}, a nonlinear method of reflection  incorporated with the inverse scattering transform has been successively applied to solve  this model (see \cite{CCAL, CZ} for the applications to other models). It relies on an extension of an initial-boundary value problem on the half-line to an initial value problem on the whole axis by imposing a special type of symmetry to the Lax pair.  Similar idea was  employed in \cite{HH1, Tarasov} where the symmetry of the Lax pair was translated into a B\"acklund transformation  connecting the space inversion symmetry,  \ie $x \to -x$, of the NLS equation. Here,  we establish  the equivalence between the nonlinear method of reflection and the inverse scattering transform for NLS on the half-line.  This is based on the extension scheme for the double-row monodromy matrix  presented in Section~\ref{sec.33}.

Consider the NLS equation on the half-line subject to Robin boundary conditions  \eqref{eq:Rb1}. We slightly modify the form of the double-row monodromy matrix \eqref{eq:drmm1} by replacing the reflection matrix $K_+$ to the left, \ie
\begin{equation}
  \label{eq:15}
  \widetilde{\Gamma}(\lambda)  = K_+(\lambda)T_\infty(t;\lambda)K_-(-\lambda)T^{-1}_\infty(t;-\lambda)\,.
\end{equation}
Clearly, this form of the double-row monodromy matrix will yield an equivalent set of scattering functions characterizing the half-line problem for NLS. Since we are dealing with the Robin boundary conditions,  the reflection matrices $K_\pm$ can be chosen as
\begin{equation}
  \label{eq:23}
   K_\pm(\lambda) = K(\lambda) =\bma 1 & 0 \\ 0& -\frac{\lambda-i\alpha}{\lambda+i\alpha}\ema\,. 
\end{equation}
Following our extension scheme \eqref{eq:qext}, using
\begin{equation}
\widetilde{T}_{-\infty}(t;\lambda) = \lim_{L\to \infty}e^{iL\lambda \sigma_3}\,\widetilde{T}_{-L}(t;\lambda)  = T_\infty(\lambda)\,,
\end{equation}
where the quantity  $\widetilde{T}_{-\infty}(t;\lambda)$ is defined in \eqref{eq:TTT}, one has 
\begin{equation}
  \label{eq:4}
  \widetilde{\Gamma}(\lambda)= K(\lambda)\widetilde{T}_{-\infty}(t;\lambda)K(-\lambda)T^{-1}_\infty(t;-\lambda)\,. 
\end{equation}

Alternatively, the nonlinear method of reflection is characterized by an extended Lax matrix on the whole axis  in  the form \cite{fokas1989initial}
\begin{equation}
  \label{eq:U22}
  \cU_{\text{ext}} =  -i\lambda\sigma_3  +\cQ_{\text{ext}}\,,
  \end{equation}
  where the extended function $\cQ_{\text{ext}}$ is defined as 
  \begin{equation}
    \label{eq:QQ28}
   \cQ_{\text{ext}}: =\cQ_{\text{ext}}(x;\lambda) = \cH(x)Q(x) - K(-\lambda)  Q(-x)\cH(-x)K(\lambda)\,,
  \end{equation}
with $Q$ being defined in \eqref{eq:laxp2} and  $\cH(x)$ representing the Heaviside step function,  \ie $\cH(x) =1$ for $x>0$ and  $\cH(x) =0$ for $x<0$. A time-part Lax matrix $\cV_{\text{ext}}$ can be similarly  defined using  $\cQ_{\text{ext}}$.  The inverse scattering transform for this extended Lax pair on the whole axis led to solutions of NLS on the half-line \cite{biondini2009solitons}, and the monodromy matrix for \eqref{eq:U22} is (which is the inverse of the full line monodromy matrix defined in \eqref{flT} by assuming vanishing boundary conditions as $x\to \pm \infty$) 
  \begin{align}
    \cT(\lambda) &= \lim_{y\to \infty} e^{-i y\lambda \sigma_3}\,\left[\overset{\curvearrowleft}{\exp}\int_{y}^{-y}  \cU_{\text{ext}}(\xi;\lambda)d\xi\right] \,  e^{-i y \lambda \sigma_3} \nonumber \\&= \lim_{y\to \infty}\left[ e^{-i y\lambda \sigma_3}\,\overset{\curvearrowleft}{\exp}\int_{0}^{-y}  \cU_{\text{ext}}(\xi;\lambda)d\xi\right] \lim_{y\to \infty}\left[ \overset{\curvearrowleft}{\exp}\int_{y}^{0}  \cU_{\text{ext}}(\xi;\lambda)d\xi\, e^{-i y \lambda \sigma_3}\right]\, \nonumber \\
    & =K(-\lambda)\,\widetilde{T}_{-\infty}(t;-\lambda)K(\lambda)T^{-1}_\infty(t;\lambda) \,,\nonumber\end{align}
which means   $\cT(-\lambda) = \widetilde{\Gamma}(\lambda)$. This implies that the scattering functions  in both cases are equivalent. Therefore, we prove the equivalence of the two approaches. Moreover, the extended  $\cQ_{\text{ext}}$ satisfies the involution relation
\begin{equation}
  \label{eq:24}
  \cQ^*_{\text{ext}}(x;\lambda^*) = \sigma_2  \,\cQ_{\text{ext}}(x;\lambda)  \,\sigma_2\,,
\end{equation}
which is compatible with that of the original Lax matrices \eqref{eq:inv1}.  This makes the nonlinear method of reflection for this particular model extremely convenient.   One has  Dirichlet boundary conditions ($\alpha_1\to 0$) with
  \begin{equation}
\cQ_{\text{ext}}(x;\lambda) = \cH(x)  Q(x) -   Q(-x)  \cH(-x)  \,, 
\end{equation}
and  Neumann boundary conditions ($\alpha_1\to \infty$) with
\begin{equation}
  \label{eq:25}
 \cQ_{\text{ext}}(x;\lambda) = \cH(x)Q(x) - \sigma_3    \,Q(-x) \cH(-x)  \,\sigma_3\,,
\end{equation}
as subcases. 

Although the nonlinear  method of reflection works well in the case of Robin boundary conditions where the reflection matrix is a constant diagonal matrix, for generic time-dependent  reflection matrix,  it is not clear how to prepare an extended $\cQ_{\text{ext}}$  in the sense of \eqref{eq:QQ28}. However, the extension scheme \eqref{eq:qext} for the half-line double-row monodromy matrix provides a precise description of the nonlinear method of reflection for the case of generic integrable boundary conditions. In this extended picture, the boundary conditions can be interpreted as defect-type conditions present at the origin. Solutions of the half-line problem are indeed solutions of the extended full line problems by restricting the space variable $x$ to the positive semi axis.  In fact,  one can argue that the extension method is not essentially necessary, since the half-line model can be solved on the half-line. 
\section{Soliton solutions on the half-line}\label{sec:5}
We provide examples of pure multi-soliton solutions of NLS on the half-line subject to the boundary conditions associated with a given $\kappa(\lambda)$ (see Fig.~(\ref{fig:h1}-\ref{fig:h4}) below).   Consider the Riemann-Hilbert problem presented in Section~\ref{sec:43}. In the special case $B(\lambda)=0$ for $\lambda\in\RR$,  a unique regular solution is the identity matrix. Singular solutions of the Riemann-Hilbert problem  can be constructed using  dressing transformations with the help of the dressing factor $D(\lambda)$  \cite{faddeev, babelon} \begin{equation}
  \label{eq:12}
  D(x,t;\lambda) = I + \frac{r_1^* -r_1}{\lambda - r_1^*}\cP(x,t)\,,\quad \cP(x,t) =\frac{\varphi_1 \varphi_1^\dagger}{\varphi_1^\dagger \varphi_1}\,, 
\end{equation}
where the superscript ${}^\dagger$ denotes the transpose conjugate, and $\varphi_1:=\varphi(x,t)$ is a particular solution of the ``undressed'' Lax pair. This form of $D(\lambda)$ characterizes a dressing factor of degree $1$, and
\begin{equation}
  \label{eq:28}
\det D(\lambda) = \frac{\lambda-r_1}{\lambda-r^*_1}\,,\quad   D(\lambda)\vert_{\lambda =r_1}\,\varphi_1 = 0\,. 
\end{equation}
The last equality  means $\varphi_1$ is a vector belonging to the kernel space of a ``zero'' of $  D(\lambda)$ evaluated at  $\lambda =r_1$. Without loss of generality, one could set $\varphi_1$ as
\begin{equation}
  \label{eq:30vp}
\varphi_1 =  \bma \gamma_1 \\ 1 \ema\,, 
\end{equation}
where $\gamma_1$ can be identified with the norming constant in the discrete soliton data. A set $\{r_j;\gamma_j\}_{j=1,\dots, N}$ forms the singular data which uniquely characterize a singular solution by iteration of dressing transformations.    

\begin{figure}[h]
  \centering
  \includegraphics[width=0.4\linewidth]{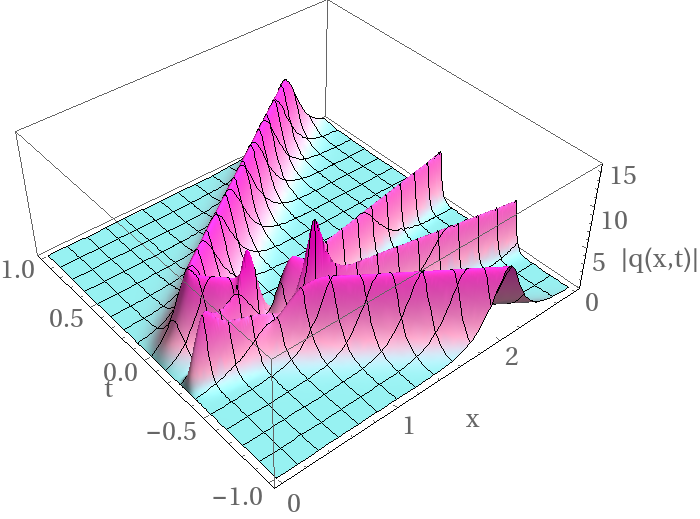} \hspace{.4cm}
  \includegraphics[width=0.3\linewidth]{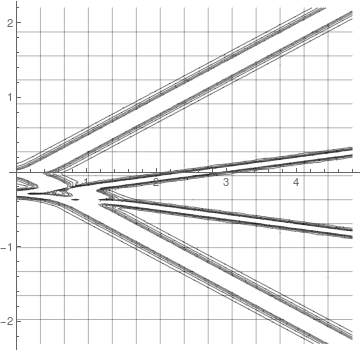}
  \caption{Two solitons on the half-line subject to Dirichlet boundary conditions. The soliton data are $\{r_1 = 0.5+     i 4, r_2 = 2+3i; \gamma_1 =4,\gamma_2 =10^6\}$ together with the paired data  $\{\widetilde{r}_1, \widetilde{r}_2; \widetilde{\gamma}_1,\widetilde{\gamma}_2\}$.}
  \label{fig:h1}
\end{figure}

Let $r_j \in \CC^{(+)}$ be complex number with nonzero imaginary part. Due to formula \eqref{eq:29SJ} (note that this formula is evaluated at $x=0$), if $r_j$ is a zero of $J^{(+)} $ and $ \varphi_j$ is a vector such that  $J^{(+)}\vert_{\lambda=r_j}\varphi_j = 0 $, then   $S^{(+)}\vert_{\lambda=r_j}\varphi_j =0$. This implies that there exists a paired vector $\widetilde{\varphi}_j$, which belongs to the kernel space of $J^{(+)}$ evaluated at $\lambda = -r_j^*$, such that
\begin{equation}
  \label{eq:20}
\widetilde{\varphi}_j =\sigma_2 K^*(r_j)\varphi^*_j\,.
\end{equation}
By setting $\widetilde{\varphi}_j$ in the  form  \eqref{eq:30vp},  one can show that the paired norming constants $\widetilde{\gamma}_j$ and $\gamma_j$  obey the relation \eqref{eq:ggr}. The kernel space associated with a pure imaginary zero of $J^{(+)}$ can be derived similarly, and the associated norming constant is characterized by \eqref{eq:12eta}. Therefore, the singular data  of  the Riemann-Hilbert problem can be identified with the discrete scattering data \eqref{eq:28dsd}. The pure soliton solutions of NLS can be reconstructed using the singular data as \cite{faddeev}
\begin{equation}
  q(x,t) = 2i \frac{\det \cR}{\det R}\,,
\end{equation}
where
\begin{equation}
  \cR = \bma  R & \begin{matrix} \gamma_1(x,t) \\ \vdots \\ \gamma_N(x,t) \end{matrix} \\ \begin{matrix} 1 &\cdots & 1 \end{matrix}  & 0 \ema \,,
  \quad R_{j k} = \frac{1 +  \gamma_k^*    \gamma_j}{r_j-r_k^*} \,, \quad \gamma_j: = \gamma_j e^{-2i \theta(x,t;r_j)}\,.\end{equation}
It suffices to employ the discrete scattering data \eqref{eq:28dsd} in the above formulae to derive pure soliton solutions on the half-line. In Fig.~\ref{fig:h1}, two soliton solutions on the half-line subject to Dirichlet are presented. The cases that  ``moving'' and ``static'' solitons coexist are shown in  Fig.~\ref{fig:h3} under the Robin boundary conditions. Fig.~\ref{fig:h4} shows examples of two soliton solutions on the half-line subject to some time-dependent boundary conditions.

\begin{figure}[h]
  \centering
  \includegraphics[width=0.4\linewidth]{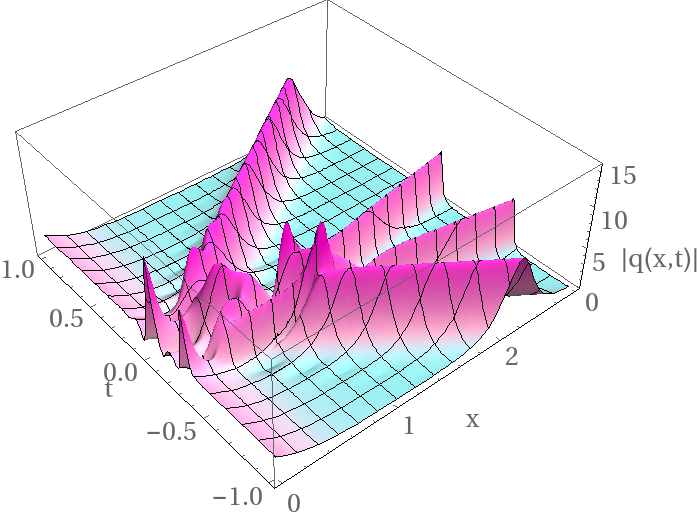} \hspace{0.2cm}
  \includegraphics[width=0.4\linewidth]{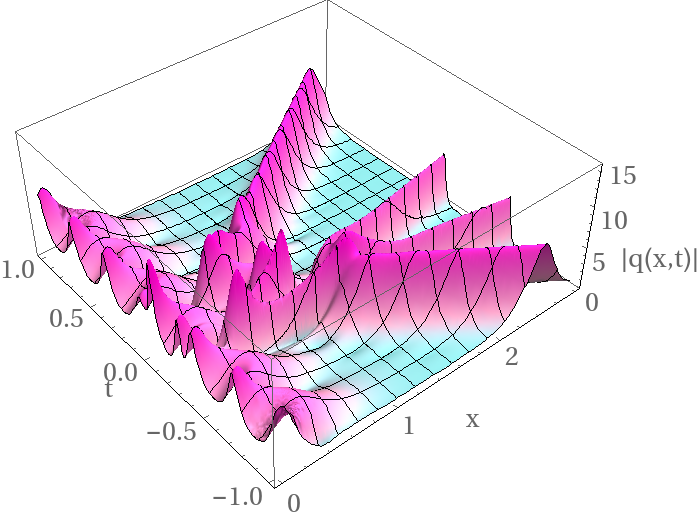}
  \caption{Robin boundary conditions ($\alpha_1=-1.5$): two moving solitons interacting with one static soliton (left) and two static solitons (right). The  moving soliton data are $\{r_1 = 0.5+     i 4, r_2 = 2+3i; \gamma_1 =4,\gamma_2 =10^6\}$ together with the paired data  $\{\widetilde{r}_1, \widetilde{r}_2; \widetilde{\gamma}_1,\widetilde{\gamma}_2\}$ for both cases. The static soliton data are respectively  $\{s_1 = 2;\eta_1 = \sqrt{7} \}$ and  $\{s_1 = 2, s_2 = 3 ;\eta_1 = \sqrt{7},\eta_2 = \sqrt{3} \}$.}
  \label{fig:h3}
\end{figure}

\begin{figure}[h]
  \centering
  \includegraphics[width=0.4\linewidth]{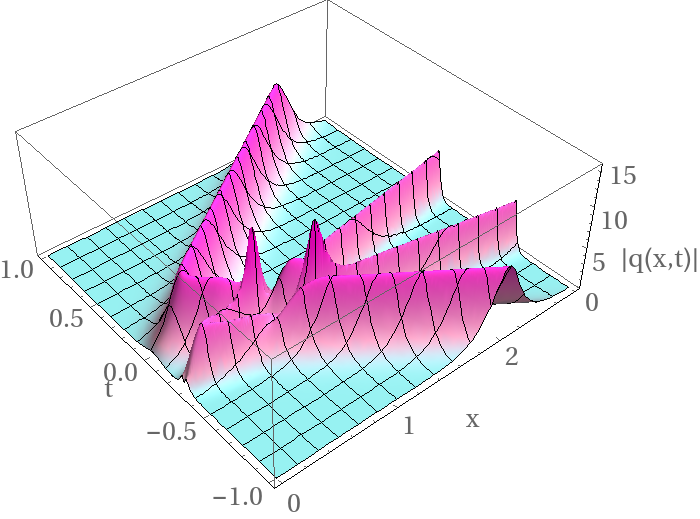} \hspace{.4cm}
  \includegraphics[width=0.4\linewidth]{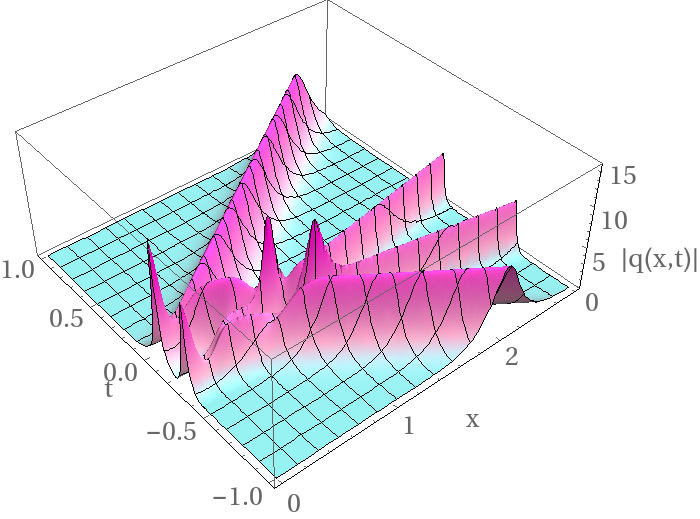}
  \caption{Two solitons on the half-line with soliton data: $\{r_1 = 0.5+     i 4, r_2 = 2+3i; \gamma_1 =4,\gamma_2 =10^6\}$ and the paired data  $\{\widetilde{r}_1, \widetilde{r}_2; \widetilde{\gamma}_1,\widetilde{\gamma}_2\}$. The boundary conditions are respectively the defect-type boundary conditions in Case $2$ with $\alpha_1 = -1.5$ (left) and the implicit boundary conditions in Case $5$ with $\beta_1 = 1 -i$ (right).}
  \label{fig:h4}
\end{figure}

\section{Conclusion}
Based on Sklyanin's formalism of integrable boundary conditions and  double-row monodromy matrix, we provide  examples of boundary conditions for the focusing NLS equation, which belong to a hierarchy of integrable boundary conditions. Some of the boundary conditions are in implicit forms involving certain additional fields coupled with the NLS fields and parameters.  We also set up the scattering system for the double-row monodromy matrix. This lays the groundwork for the direct scattering transform for the interval problems for the NLS equation.  The  direct scattering transform for NLS on the half-line is established by extending one endpoint of the interval to infinity, and the inverse part is formulated as a Riemann-Hilbert problem. As a particular application, multi-soliton solutions of NLS on the half-line are computed.

The present work provides an analytic method for solving initial-boundary value problems for NLS on the  half-line subject to a class of boundary conditions. In contrast to Fokas' unified transform method, integrable boundary conditions are required in our approach. The nonlinear method of reflection can be seen as an equivalent approach using a systematic extension scheme provided in Section \ref{sec.33}.  This suggests that the solution structures of initial-boundary value problems of NLS  on the half-line with integrable boundary conditions are essentially the same as those of initial value problems of NLS on the whole axis.  In both cases, the spectral parameter is living  on the complex plane, and the jump conditions of the Riemann-Hilbert problem are defined on the real axis.    Although we only consider the NLS model as a particular example, the  inverse scattering transform for half-line problems  we present in this paper can be readily generalized to a wide range of integrable PDEs. This includes models such as  the sine-Gordon equation, the Ablowitz-Ladik equation and the (focusing or defocusing) NLS on a constant background, etc. A natural and important continuation of the present work is tackle  initial-boundary value problems of  NLS on an interval equipped with integrable boundary conditions. It is expected that  the solution structures of interval problems would be similar to those of  periodic problems for NLS.


\appendix
\section{Semi-discrete Lax pair formulation  for $K_+(\lambda)$ }\label{ap:0}
Taking $K_+(\lambda)$ in the form \eqref{eq:KETA}, we are looking for the existence of a column vector $\phi(\lambda)$ satisfying
\begin{equation}
  \label{eq:13AA}
  \bma 1 & 0 \\ 0 & \kappa(\lambda) \ema    \phi(\lambda) = \eta\, \phi(-\lambda)\,.
\end{equation}
Choose $\eta = \kappa(\lambda)$, and let  $\phi(\lambda)$ be in the form
  \begin{equation}\phi(\lambda) = \bma v(\lambda) \\ 1\ema\,. \end{equation}
Then, \eqref{eq:13AA} is reduced to
\begin{equation}
  \label{eq:8v}
  v(\lambda) = \kappa(\lambda)v(-\lambda)\,.
\end{equation}
Taking the form  of $\kappa(\lambda)$ as $\kappa(\lambda) = \epsilon f(\lambda) /g(\lambda)$ with $\epsilon = \pm 1$, {\it cf}.~\eqref{eq:normk1},  $f(\lambda)$ and $g(\lambda)$ are connected by
  \begin{equation}
    \label{eq:22}
    f(\lambda) =  (-1)^\cM g(-\lambda)\,.
  \end{equation}
  If the quantity $\epsilon (-1)^\cM = 1$,  then \eqref{eq:8v} becomes
  \begin{equation}
    \label{eq:18}
      v(\lambda) = \frac{f(\lambda)}{f(-\lambda)}v(-\lambda)\,.
  \end{equation}
  It suffices to choose $v(\lambda)  = f(\lambda)$. If the quantity  $\epsilon (-1)^\cM = -1$,  then \eqref{eq:8v} becomes
  \begin{equation}
    \label{eq:13}
  v(\lambda) = -\frac{f(\lambda)}{f(-\lambda)}v(-\lambda)\,.
  \end{equation}
In this case,  one could set $v(\lambda)^2  = f(\lambda)^2$, or $v(\lambda)  = \sqrt{f(\lambda)^2}$ by fixing the branch of the square root with the sign of $\lambda$. 

\section{Proof of Lemma \eqref{lemma:1}}\label{ap:1}
It was shown in Section~\ref{sec:42} that the element $A(\lambda)$ (resp. $A^*(\lambda)$) can be analytically extended to the upper half (resp. lower half) complex plane. Consider the equation  \eqref{eq:AA1}   as a multiplicative Riemann-Hilbert problem. By taking the logarithm,  it is transformed to an additive Riemann-Hilbert problem 
\begin{equation}  \label{eq:laa}
\log A(\lambda) +   \log A^*(\lambda) =\log (1-|B(\lambda)|^2 ) \,,\quad \lambda\in \RR\,,  
\end{equation}
with the asymptotic behaviors
\begin{equation}
  \label{eq:laa1}
\lim_{\lambda\to \infty} \log A(\lambda)= \lim_{\lambda\to \infty} \log A^*(\lambda)= 0\,, 
\end{equation}
which is a result of \eqref{eq:21AA}. 
We assume  $B(\lambda)$ to be  a smooth function. This is justified using the fact that the initial data $q(x,t_0)$ are smooth. Moreover,  $|B(\lambda)|<1$ for $\lambda\in \RR$. This follows from the assumption that $A(\lambda)$ has finitely many simple zeros strictly located in $\CC^{(+)}$.   $B(\lambda)$ also satisfies the extra  involution relation 
\begin{equation}\label{eq:aafbb11}
  \kappa(\lambda)B(\lambda) = -B(-\lambda)\,,
\end{equation}
where  $\kappa(\lambda)$ is defined  in Section~$3.2$. 
The above Riemann-Hilbert problem \eqref{eq:laa} with the normalization conditions \eqref{eq:laa1} has a unique regular solution (see, for instance, \cite{AFF}). Let
\begin{equation}
  \label{eq:8}
A(\lambda) = \zeta(\lambda)\widetilde{A}(\lambda)\,,
\end{equation}
where $\widetilde{A}(\lambda)$ be the regular solution. Applying the Cauchy integral operator to \eqref{eq:laa}, one has $\log \widetilde{A}$ in the form
\begin{equation}
  \label{eq:tAB}
  \log \widetilde{A}(\lambda) = \frac{1}{2 i \pi }\int^{\infty}_{-\infty} \frac{\log (1-|B(\mu)|^2 )}{\mu-\lambda}d\mu\,,
\end{equation}
for $\lambda \in {\CC}^{(+)}$. This quantity can be extended to the real axis using the Sokhotski-Plemelj formula. By taking account of the  relation \eqref{eq:aafbb11}, one has
\begin{equation}\label{eq:BBFBB}
 |B(\lambda)|^2 =-\kappa(\lambda)B(\lambda)B^*(-\lambda) =|B(-\lambda)|^2\,,\quad \lambda \in \RR\,.
\end{equation}
Then, $\log \widetilde{A}(\lambda)$ is reduced to \begin{equation}
  \label{eq:11}
  \log \widetilde{A}(\lambda) = \frac{1}{ 2i \pi }\left(\int^{\infty}_{0}+\int^{0}_{-\infty}\right) \frac{\log (1-|B(\mu)|^2 )}{\mu-\lambda}d\mu =\frac{1}{ i \pi }\int^{\infty}_{0} \lambda\frac{\log (1-|B(\mu)|^2 )}{\mu^2-\lambda^2}d\mu\,. 
\end{equation}
By adding the singularities, one gets the desired results.

\end{document}